\documentclass[aps,reprint,showpacs,superscriptaddress,
footinbib,citeautoscript]{revtex4-1}

\usepackage{graphicx}

\usepackage[utf8]{inputenc}
\usepackage{csquotes}
\usepackage[american]{babel}
\usepackage[T1]{fontenc}
\usepackage{enumerate}
\usepackage{mdwlist}
\usepackage[activate=normal]{pdfcprot}
\usepackage{bbding}
\usepackage{color}
\usepackage{amssymb}
\usepackage{amsmath}
\usepackage{amsfonts}
\usepackage{mathrsfs}

\usepackage{bm}
\usepackage{dcolumn}
\usepackage{color}
\usepackage[colorlinks=true,citecolor=blue]{hyperref}
\hypersetup{colorlinks=true,citecolor=blue,linkcolor=red
,urlcolor=blue}

\usepackage{amsmath,amssymb}
\usepackage{color}
\usepackage{graphicx}
\usepackage{bbold}
\usepackage{feynmf}
\usepackage{natbib}

\def\k{{\bf k}}
\def\q{{\bf q}}
\def\G{{\hat{\bf G}}}
\def\Sig{{\hat{\bf \Sigma}}}

\def\F{{\bf F}}
\begin{document}
\title{Superconducting critical temperature of hole doped blue phosphorene}
\author{Davoud Nasr Esfahani}
\affiliation{Condensed Matter National Laboratory, Institute for Research in Fundamental Sciences (IPM), Tehran 19395-5531, Iran}
\author{Reza Asgari}
\affiliation{School of Physics, Institute for Research in Fundamental Sciences (IPM), Tehran 19395-5531, Iran}
\affiliation{School of Nano Science, Institute for Research in Fundamental Sciences (IPM), Tehran 19395-5531, Iran}
\date{\today}

\begin{abstract}
We theoretically explore the superconducting critical temperature of hole doped blue phosphorene. Implementing the density functional theory calculations, we show that for the hole doped blue phosphorene, the isotropic superconducting state is induced owing to the quite
strong electron-phonon coupling.
The theory is based on the Migdal-Eliashberg formalism and the critical temperature
is obtained through set-of-equations, self-consistency. In addition, we include a vertex correction diagram
to the Migdal-Eliashberg formalism. The inclusion of the vertex correction beyond the Migdal-Eliashberg formalism changes the $T_c$ about $\pm20$K, depending on the level of the doping. Our accurate numerical results show that the superconducting critical temperature
is still quite high, even in the cases that the vertex correction is implemented.
\end{abstract}

\maketitle

\section{Introduction}
Two-dimensional (2D) superconductivity has attracted much attention for the past decade and its explore has provided insight into a variety of rich physics occurs at the level of quantum phenomena.
The fabrication of monolayer cuprate superconductors opens a new venue to investigate 2D materials and afterwards many new fabricated techniques such as molecular beam epitaxy together with the surface
reconstruction process, mechanical exfoliation, and different methods for the production of field effect devices were introduced into the field of 2D supercondutors~\cite{2Dsuper}. Nowadays, researches
can access superconductivity at the 2D limit in new advanced 2D crystalline materials.

In a 2D material with $N$ layers, the Bardeen-Cooper-Schrieffer (BCS) theory~\cite{BCS} predicts that $k_BT_c(N) = 1.13E_D \exp(-1/(UN(\varepsilon_{\rm F})N))$
where $N(\varepsilon_{\rm F})$ is the single layer density of states (DOS) at the Fermi energy,
$U$ is the pairing interaction strength and $E_D$ is the Debay cutoff energy. Moreover, the critical field in a strictly 2D BCS superconductor is the Pauli paramagnetic limit,
$H_p=\Delta/(\sqrt{2}\mu_B)$ where $\mu_B$ is the Bohr magneton and $\Delta$ is the cooper pair energy gap. Above the
Pauli field $H_p$, the Zeeman splitting of the Cooper pairs compensates
the energy gained from creating the BCS condensate and therefore,
2D superconductivity is suppressed.

Black phosphorene (BP), a single layer analog of black phosphuros with puckered structure,
has recently been exfoliated \cite{Liu}. Tempted by successful synthesis of BP, several other monolayer structures have been
proposed for phosphorus allotropes~\cite{Wu,Soares}. Among them,
blue phosphorene (BLP), which is a semiconductor with a buckled honeycomb structure and energy gap $\sim 2 eV$, is energetically the most stable one after monolayer BP \cite{Wu}. Moreover,
it was shown that BLP is dynamically stable \cite{Zhu} and thermodynamically is more stable than BP in elevated temperatures \cite{Aierken}. Furthermore, BLP is recently realized through epitaxial growth~\cite{Zeng}.
In terms of applications, the BP has been proposed as a high mobility material appropriate for a conventional field effect transistor applications \cite{Liu}. On the other hand,
it has been shown that the value of the mobility could be much smaller, in particular, owing to the anisotropy in the material and larger phase-space for the electron-phonon coupling (EPC) \cite{Liao}.
Later on, an EPC-mediated critical superconducting temperature $T_c\sim 17 K$ was reported for electron doped BP by
using Lithium adsorption \cite{Sanna}. Its intercalation by several alkali metals (Li, K, Rb and Cs) has been described
recently~\cite{Geim} and all the intercalated compounds have been found to be
superconducting, exhibiting the same critical temperature of $3.8 \pm 0.1$K and practically identical characteristics
in the superconducting state~\cite{Geim}. Furthermore, a superconducting temperature above 20K was recently predicted
 for electron doped bilayer-BLP through intercalation by alkali metals and alkaline earths.\cite{jun}

Early proposal for superconducting state in 2D materials refers to the electron doped graphene where the critical superconducting temperature
is $T_c\sim 15 K$ and the electron doping was realized by Lithium adsorption on graphene \cite{natphys, mauri2, Ludbrook}.
However, recent experiments proposed new way of graphene superconductivity by activating the dormant potential for graphene  in its own right by coupling it
with a material called praseodymium cerium copper oxide~\cite{bernardo}. Besides, 2D systems showing an electric-field-induced superconductivity~\cite{Ueno}.

In pristine graphene, there are at least two features which are detrimental to superconductivity at arbitrary low temperatures. First of all, the presence of the zero DOS at the Fermi level and second, the occurrence of a horizontal
mirror plane ($\sigma_h$ symmetry) in graphene. While the former is a prerequisite for the superconductivity, the latter only suppresses the EPC,
i.e. the linear coupling to flexural modes is forbidden due to the symmetry considerations (the presence $\sigma_h$ symmetry) \cite{Manes, Fischetti}. The role of the lithium adsorption on graphene is two folded. First, its role is to make a finite DOS at the Fermi energy and second, it promotes the coupling of the electrons to flexural modes which
leads to an enhanced the EPC \cite{Boeri,mauri1,mauri2}.

Compared to graphene, since pristine BLP is an insulators, the electron/hole doping is necessary to induce a critical superconducting state.
The required doping could be achieved by the electrical doping or by dopant or ad-atoms~\cite{bblp}. On the other hand, BLP has an inversion symmetry and $\sigma_h$ does not hold, hence, it is expected that
phonons with out-of-plane distortions are intrinsically coupled to the electrons and therefore no ad-atoms are necessary to promote
the coupling to the out-of-plane distortions. The essence of the out-of-plane distortions could be understood by the fact that the majority of the states near the valence band maximum has $p_z$ character.

In this paper, we carry out first-principles calculations to calculate the electron-phonon coupling of BLP to investigate the superconducting features of the system. Our theory is stemming from the multi-band Migdal-Eliashberg \cite{migdal,eliashberg} theory including the second-order self-energy as a vertex correction.
Our numerical results show that a higher superconducting critical temperature occurs at lower hole density and thus the critical temperature ranging from $100$ to $40$K are obtained by considering the hole
densities between $5\times10^{13}$ to $3.8\times 10^{14}$ cm$^{-2}$.

This paper is organized as follows.
In Sec.~\ref{sec:model}, we present the methodology used to calculate the superconducting state in the system and also describe second-order self-energy as a vertex correction. A set-of-equations is solved self-consistency to calculate the energy gap and critical temperature of the system. In Sec.~\ref{sec:result} we present and describe main results of the superconducting state in the system and finally, we conclude and summarize our main results in Sec.~\ref{sec:conc}

\section{Theory and DFT computational simulations}\label{sec:model}

In order to compute the electronic and phononic band dispersions of the system, the density functional theory (DFT) and density
functional perturbation theory (DFPT) \cite{baroni}, as implemented in {\it Quantum Espresso}~\cite{esspreso},
are employed. The generalized gradient
approximation in the scheme of Perdew, Burke, and
Ernzerhof \cite{gga} and norm-conserving pseudopotentials
are used throughout our calculations. Moreover, the Wannier interpolation of quantities (such as electron dispersions,
phonon dispersions and electron-phonon interactions) from a fully self-consistent calculations on a coarse mesh to a fine mesh is applied
as implemented in {\it EPW} code~\cite{Giustino1,Giustino2}, which is an integrated code into Quantum Espresso.
In order to avoid interactions between layers along the $z$ direction, a vacuum of 20~\AA between layers is considered.
Most of the results are examined between calculations within two different parameters set in some instances throughout this paper to provide increased precision for critical results and convergence check.
These sets are namely DFT self-consistent calculations with integration over $12\times 12$ $k-$ Monkhrost-Pack mesh, plane-wave energy cutoff of $70$ Ry followed by
 DFPT calculation on a $10 \times 10$ $q-$mesh. The interpolation is performed on a $10\times 10$ (uniform coarse electronic) $k-$mesh and $10\times 10$  (uniform coarse phononic)
$q-$mesh, and the second parameters set consists of the DFT self-consistent calculations with integration over $20\times 20$ $k-$ Monkhrost-Pack, plane-wave energy cutoff of $90$ Ry followed by  DFPT calculation on a $16 \times 16$ $q-$mesh. The interpolation is performed on a $16 \times 16$ (uniform coarse electronic) $k-$mesh and $16\times 16$ (uniform coarse phononic) $q-$mesh. No significant deviations between the two cases is observed. The results presented in the following are based on the latter set of parameters.
Moreover, the Wannierisation (spin unpolarized) is performed using Wannier90 code~\cite{wannier} for the first $8$
bands with an initial projection into S and P orbitals, and are checked with initial projections into SP$^3$.
Furthermore, disentanglement procedure is employed and an upper bond for an inner window equal to $4.6$ eV above the valance band maximum (VBM) is set, an average spread $\sim$ 2.18 \AA$^2$/per orbital was achieved. No imaginary part is observed for the real space representation of the resulting Hamiltonian.

\begin{figure}[h]%
 \includegraphics*[width=0.3 \textwidth]{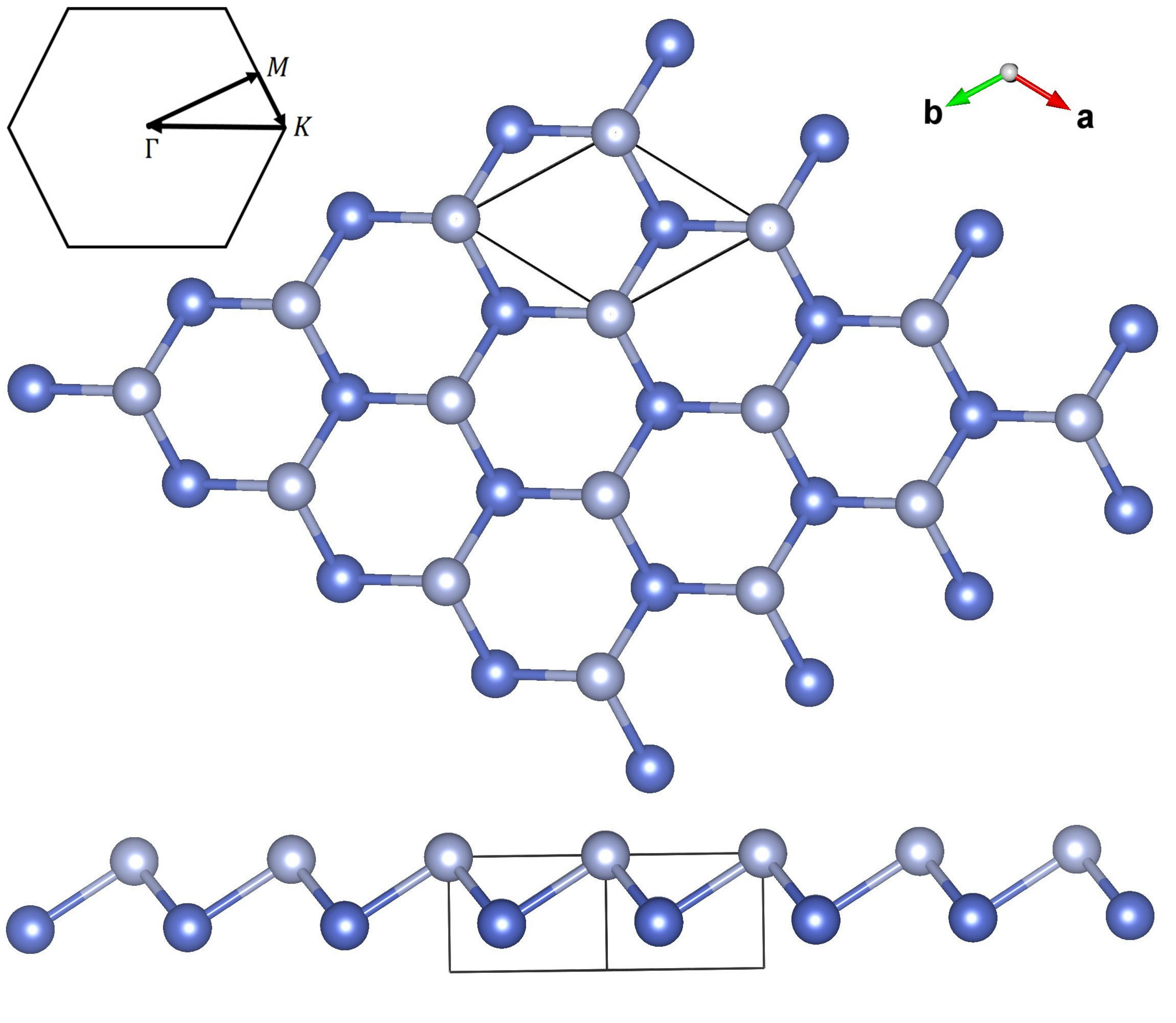}
  \includegraphics*[width=0.4 \textwidth]{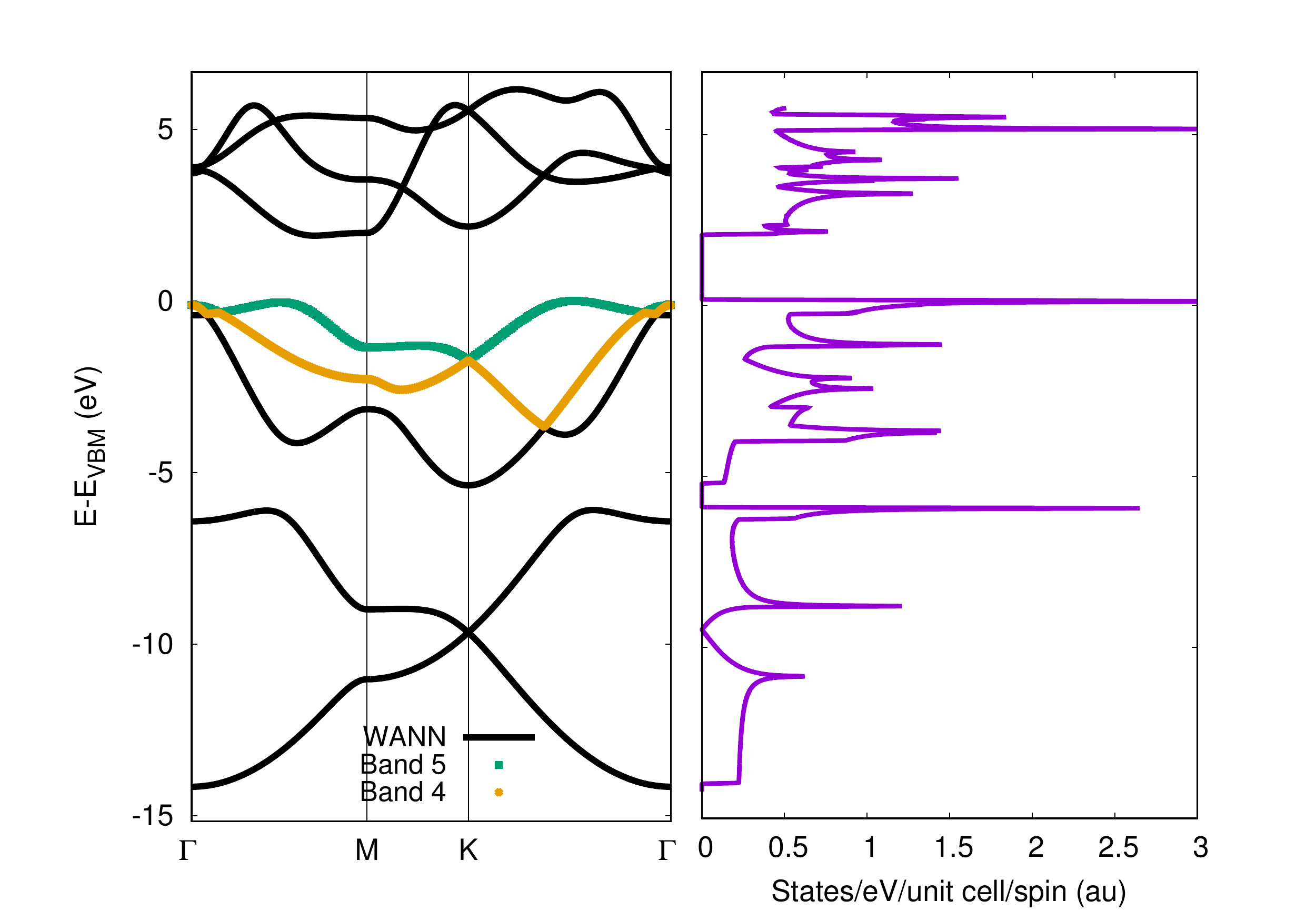}%
  \caption[]{(Color online) The top and side views of the unit-cell of BLP and the $k$-path through high-symmetry points of the Brillouin zone. The band structure and DOS of BLP calculated within two different simulations, namely self-consistent calculation and Wannier interpolated
     band structure and they are essentially the same. An indirect bang gap around $1.9$ eV is obtained. Two bands in the valence region indicate as band $4$ and $5$ for further purposes. Importantly, a flat feature of the band structure near the VBM results in a van Hove like peak at the vicinity of the VBM where $E-E_{VBN}=-0.055$ eV.}
    \label{dosband}
\end{figure}

We calculate the band structure and electronic DOS of BLP within two different approaches, i.e. a fully self-consistent
calculation and Wannier interpolated bands. Our numerical results show that those results are in very good agreement
and thus the results within the Wannier interpolation method are shown in Fig.~\ref{dosband}. An indirect band gap about $1.9$ eV is obtained between the VBM and the conduction band minimum (CBM) within the DFT simulations. In the following, we investigate BLP under the rigid shift of the Fermi energy deep into the lower energies below the VBM. Owing to the presence of a flatten band right near the VBM, the DOS shows a sharp peak.
Notice that there is a discrepancy between the electronic band structure of BLP with that of monolayer phosphorene~\cite{asgari}. For the sake of completeness, in Fig.~\ref{dosband} we indicate two valence
bands as band 4 and 5 labeled with red and green, respectively, for later purposes in order to understand the contribution of those bands on a superconducting quantity, $\alpha^2{\rm F}$.

\begin{figure}[h]%
  \includegraphics*[width=0.4\textwidth]{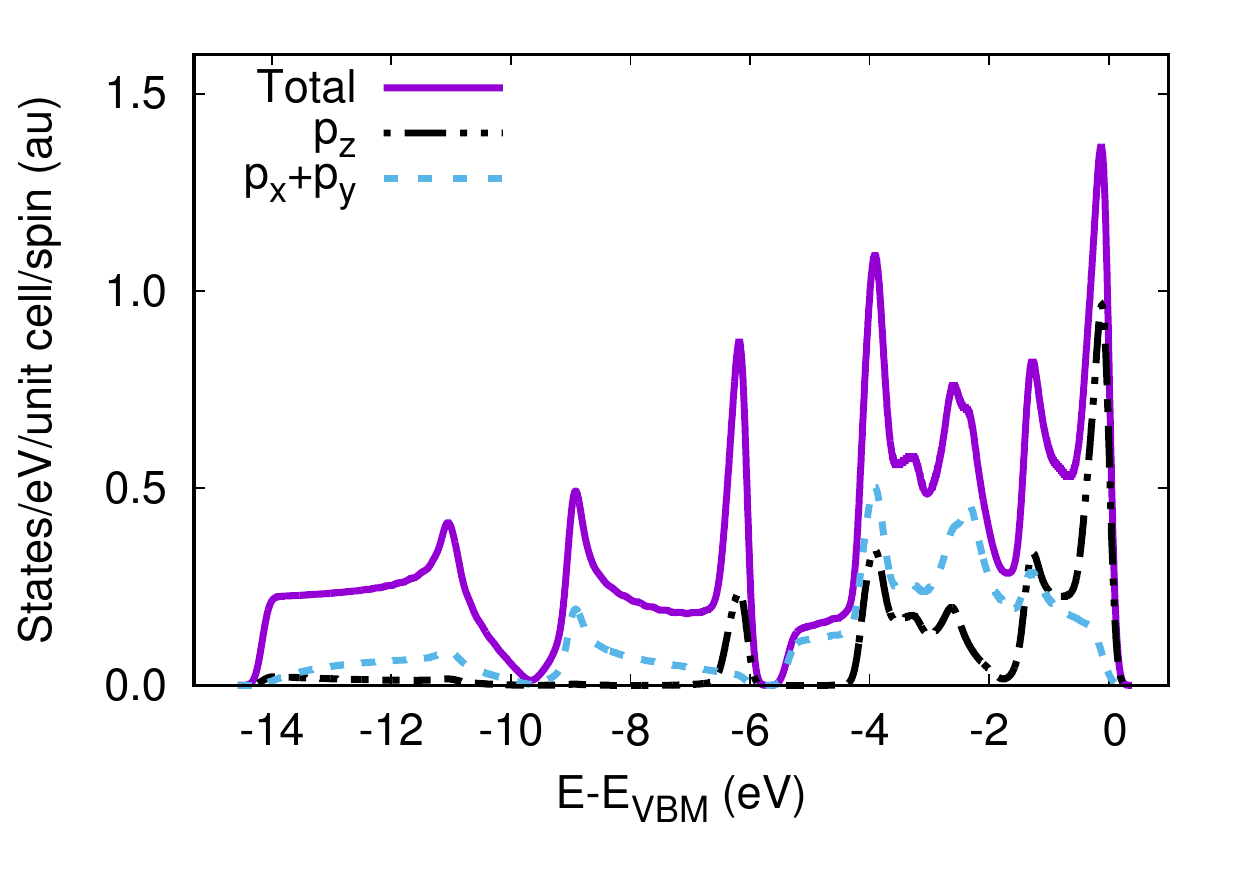}%
  \caption[h]{%
    (Color online) Projected DOS, the states near
     the VBM illustrates $p_z$ character, however, deeper into the valence band originates mainly from $p_x+p_y$. }
    \label{pdos}
\end{figure}

In Fig.~\ref{pdos}, the projected DOS into $p_x$, $p_y$ and $p_z$ is presented. Projected DOS is calculated
on a Monkhrost-Pack $20\times 20$ $k-$mesh.
Importantly enough, the states near the VBM has $p_z$ character.
Deeper into the energies and lower than the VBM, the flatness of the bands is suppressed which results in a sharp reduction of the DOS (see Fig.~\ref{dosband}). This reduction in the DOS is mostly owing to the suppression
of $p_z$ states, where the contribution of the $p_x+p_y$ is enhanced.

\begin{figure}[h]%
  \includegraphics*[width=0.4\textwidth]{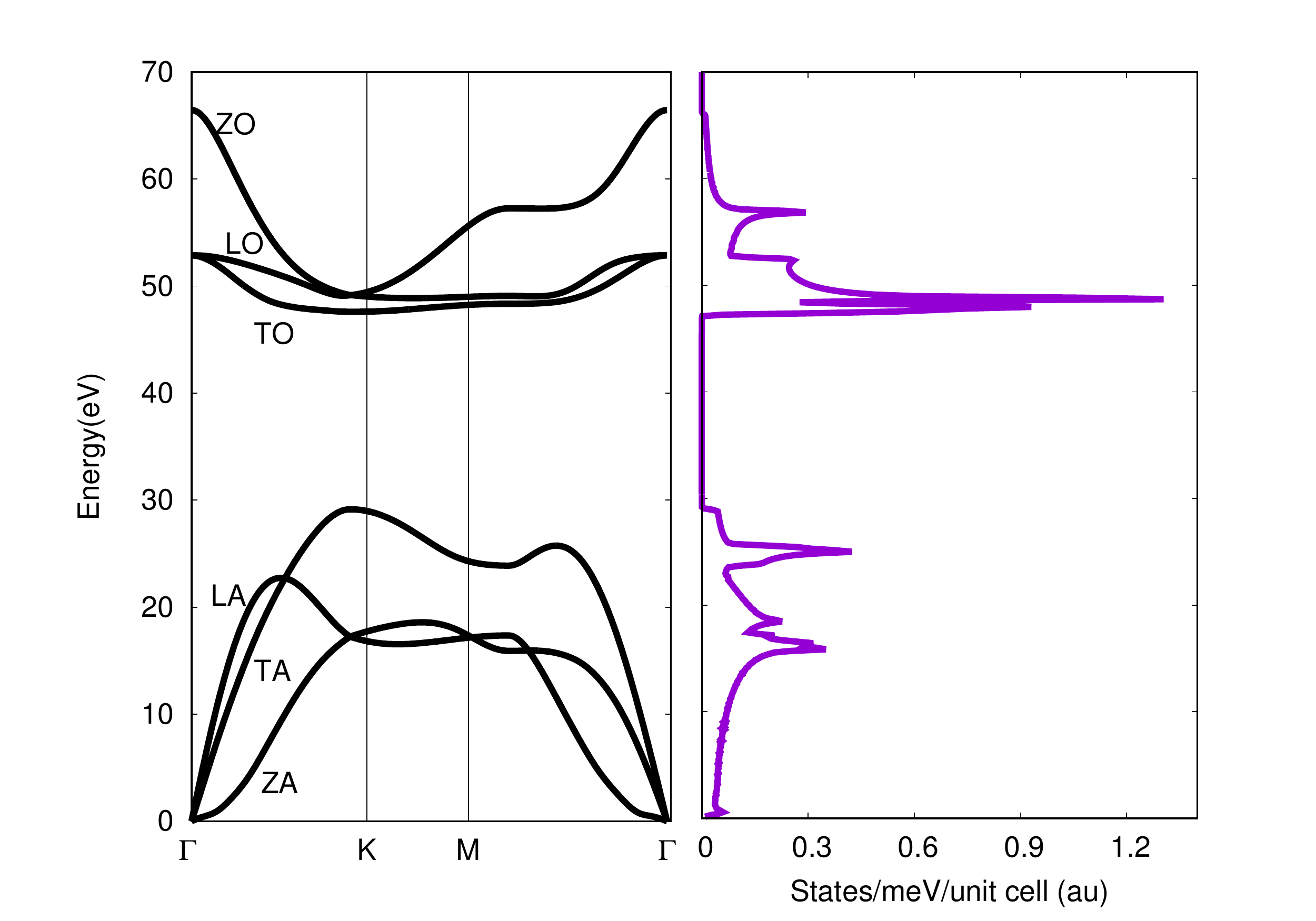}%
  \caption[h]{%
   (Color online) The dispersion of the phonons and corresponding modes of BLP. In the low-energy, there are three different bands, two of these are characterized by the
in-plane displacements (acoustic modes) at longer phonon wave lengths. The other remaining mode has major out-of-plane
displacement at longer wavelengths. This mode is softer than the other two modes and for a perfectly planar 2D material its energy dispersion acquires a $\sim  q^2$ dispersion relation.}
    \label{phdosband}
\end{figure}

In Fig.~\ref{phdosband} the phonon dispersion and phonon DOS of BLP are shown.
The low-energy phonon modes are composed of three different bands, where two of them are characterized by the
in-plane displacements at longer phonon wave lengths which are marked by longitudinal acoustic (LA) and transversal acoustic (TA) phonon modes,
respectively. These modes acquire linear dispersion at longer wavelengths with sound velocities. The other remaining mode has major out-of-plane
displacement at longer wavelengths which is marked by ZA. This mode is softer than the other two modes and for a perfectly planar 2D material its energy dispersion acquires a $\omega_{\q}\sim q^2$ relation. It is worth mentioning that in the BP the sound velocities in the $\Gamma-Y$ direction calculated as $7.59$ km/s and $4.48$ km/s
for LA and TA modes, respectively~\cite{asgari}.
Along the $\Gamma-X$ axis, on the other hand, the sound velocities obtained as $5.69$ km/s and $5.27$ km/s
for longitudinal and transverse vibrations, respectively~\cite{asgari}. However, BLP acquires
 almost isotropic sound velocities along the $\Gamma-M$ and $\Gamma-K$ directions. The sound velocities in BLP are $8.3$ km/s and $5.5$ km/s for the longitudinal and transversal atomic motions, respectively. The longitudinal mode has a slightly greater velocity with respect to one reported in ~[\onlinecite{Zhu}]. Moreover, the ZA mode in BP is
different with respect of the BLP.

At the same time that the ZA modes have out-of-plane displacements for long wavelengths, however, for the BLP, the ZA mode acquires a tiny in-plane displacement as well, owing to its buckled nature. This small mixture between the in-plane and out-of-plane displacement results in a $\beta_1 q + \beta_2 q^2$ dispersion for ZA mode at the long-wave lengths, where $\beta_1$ is very tiny. In particular, $\beta_1$ is very sensitive to the values of the energy cutoff and the method used for imposing the acoustic sum rule~\cite{asr}. The higher energy phonons are composed of three branches of the out-of-phase displacements and are marked by transversal optical (TO), longitudinal optical (LO)
modes for the modes with major in-plane displacements and ZO for the modes with major out-of-plane displacement.
These two groups, i.e. optical and acoustic modes,
are separated by a gap equal to $\sim 17$ meV in phononic spectrum. Comparing the phonon dispersion with the phononic DOS in Fig.~\ref{phdosband}, we see that there are sharp peaks in the phononic DOS for which
the dispersion shows flatten features as a function of phonon wavevector.

\subsection{Theory of superconductivity and Model}

In normal state, the Landau Fermi liquid theory appears to work well. The Coulomb interactions give rise to a well-defined quasiparticle
with a proper energy dispersion near the Fermi surface and they are assumed to exist.
However, the phase transition to the superconducting state invalidates the perturbation approach. Nambu~\cite{Nambu} showed how the formalism used in the normal state can be rewritten in
such a way that the diagrams used to deal with the normal state are applicable for superconducting state.

In a system with $N$ separated bands, one may extend the two-component spinor of the Nambu formalism to a $2N$ component spinor and write the total Hamiltonian in the basis.
To commence with, we first drop the Coulomb interaction in the Hamiltonian and consider a system incorporates the itinerant electrons, phonons and the electron-phonon interactions. The reason to do so, is to explore the impact of
different bands which are very close to the edge of the valence band maximum (see Fig.~1). Afterwards, we add the Coulomb interaction and many-body self-energies in the model. The spinor of the Nambu formalism is
 \begin{equation}
  \psi_\k=
\left (
  \begin{tabular}{c}
   \vdots \\
   $c_{\k i\uparrow}$ \\
  $c^{\dagger}_{-\k i\downarrow}$  \\
   \vdots \\
  \end{tabular}
\right )
 \end{equation}
where $i=1\cdots N$ is the band index, $c^{\dagger}_{\k i\sigma}$ ($c_{\k i\sigma}$) is the creation (annihilation) operator for an electron in the band $i$, reciprocal vector $\k$ and spin $\sigma=\uparrow,\downarrow$. The Hamiltonian of the system is given by,
 \begin{widetext}
 \begin{eqnarray}
  \hat{\cal {H}}_0 =&& \sum_\k \psi^\dagger_{\bf k} \hat{\varepsilon}_\k \hat{S}\psi_{\k} + \sum_{\q,\nu} \omega_{\q\nu}b^\dagger_{\q,\nu}b_{\q,\nu} +
  \sum_{\k \q \nu} (b^\dagger_{\k-\k',\nu}+b_{\k'-\k,\nu}) \psi^\dagger_{\k'}\hat{S} \hat{g}^\nu_{{\bf k'},{\bf k}}\psi_{\bf k}
  \end{eqnarray}
  \end{widetext}
where $\hat{S}$, $\hat{g}^\nu_{{\bf k'},{\bf k}}$ and $\hat{\varepsilon}_\k$ are $2N \times 2N$
matrices which elements of the $\hat{S}$ satisfies $\hat{S}_{ij} = (-1)^{i-1}\delta_{ij}$,
  elements of the matrices $\hat{g}^\nu_{{\bf k'},{\bf k}}$ and $\hat{\varepsilon}_\k$ satisfy the following relations,
 \begin{eqnarray}
&&\displaystyle {[{\hat{g}^{\nu}_{ \bf k,k'}}]}_{2i-1,2j-1}= g^{\nu\uparrow}_{ \k i, \k'j},~~~~~~ {[{\hat{g}^{\nu}}_{ \k, \k' }]}_{2i,2j} =\displaystyle g^{\nu\downarrow}_{-{\bf k}i,-{\bf k'}j} \nonumber \\
&&\displaystyle {[{\hat{g}^{\nu}}_{ \k,\k'}]}_{2i,2j-1}= 0,~~~~~~~~  {[\hat{g}^{\nu}_{ \k, \k' }]}_{2i-1,2j} = 0 \nonumber \\
&&\displaystyle{[\hat{\varepsilon}_{\bf k}]}_{2i-1,2j-1} = \delta_{ij}\varepsilon_{{\bf k} i\uparrow} ,~~~~~~ {[\hat{\varepsilon}_{\bf k}]}_{2i,2j} = \delta_{ij}\varepsilon_{{-\bf k}i\downarrow} \nonumber \\
&&\displaystyle [\hat{\varepsilon}_{\bf k}]_{2i-1,2j}= 0, ~~~~~~~~ [\hat{\varepsilon}_{\bf k}]_{2i,2j-1} = 0
 \end{eqnarray}

Here, $i$ and $j$ are band indexes, $\hat{\varepsilon}_{\bf k}$ is the single-electron block energy relative to the Fermi level, with $\varepsilon_{{\bf k}i\sigma}=\tilde{\varepsilon}_{{\bf k}i\sigma}- E_{\rm F}$, where $\tilde{\varepsilon}_{{\bf k}i\sigma}$  is the energy dispersion which is extracted from DFT calculations.
In the following we will represent the Fermi energy shift ($\delta E_{\rm F}$) corresponding to the VBM in the band structure calculation such that $E_{\rm F}=E_{VBM}+\delta E_{\rm F}$.
Notice, $\delta E_{\rm F}$ is an input parameter which we use it for a rigid scan of the band structure and within the above mentioned definition of $\varepsilon_{{\bf k}i\sigma}$
the chemical potential is readily set to zero. $\hbar \omega_{{\q}, \nu}$ is the phonon energy of the wave vector $\q$ and mode $\nu$ and $g^{\nu\sigma}_{\k i,\k' j}$ is electron-phonon matrix element (for the accurate definition of $g^{\nu\sigma}_{\k i,\k' j}$ see Appendix.~\ref{sec:appa}).

By imposing the time reversal symmetry, we thus have
$\displaystyle{g^{\nu\uparrow}}_{ {\bf k}i, {\bf k'}j} = {g^{\nu\downarrow}}_{-{\bf k}i,-{\bf k'}j}={g^{\nu}}_{{\bf k}i,{\bf k'}j}$
 and $\varepsilon_{\k i\uparrow}=\varepsilon_{-\k i\downarrow}=\varepsilon_{\k i}$. Therefore, hereafter, we drop the spin index for the band dispersions and the electron-phonon couplings. The single-particle electronic Green's function is now a $2N \times 2N$ matrix
 \begin{equation}
  \hat{G}({\bf k},\tau) = -\langle {\cal {T}}[ \psi_{ \bf k}(0){\psi_{\bf k}}^\dagger(\tau)]\rangle
 \end{equation}
which its elements read as,
 \begin{eqnarray}
 \displaystyle[\hat{G}({\k},\tau)]_{2i-1,2j-1} &=-& \langle {\cal {T}}[ c_{\k i\uparrow}(\tau) {c^\dagger}_{\k j\uparrow}(0)]\rangle,\nonumber \\
\displaystyle [\hat{G}(\k,\tau)]_{2i,2j} &= -& \langle {\cal {T}}[ c_{-{\bf k}i\downarrow}(\tau)  {c^\dagger}_{-{\bf k}j\downarrow} (0)]\rangle \nonumber\\
\displaystyle   {[\hat{G}(\k,\tau)]}_{2i-1,2j} &= -& \langle {\cal {T}}[ c_{\k i\uparrow}(\tau)  c_{-\k j\downarrow}(0)]\rangle , \nonumber\nonumber \\
 \displaystyle {[\hat{G}({\bf k},\tau)]}_{2i,2j-1} &=-& \langle {\cal {T}}[ {{c^\dagger}_{-{\bf k}j\downarrow}(\tau)c^\dagger}_{{\bf k}i\uparrow}(0) ]\rangle
\end{eqnarray}
where $\cal {T}$ is time ordering on the imaginary time axis with
$-\beta<\tau<\beta$ where $\beta$ is the inverse of temperature ($k_B=1$) and $\langle ....\rangle$ is grand canonical average.
The Fourier components of the $\G$ and $D$, where $D$ refers to the single particle phonon Green's function, are expressed as
 \begin{eqnarray}
  D_{\nu}(\q,\tau) = \frac{1}{\beta}\sum_{n=-\infty}^{\infty}e^{-i\nu_n\tau}D_{\nu}(\q,i\nu_n)\nonumber\\
  \G(\k,\tau) = \frac{1}{\beta}\sum_{n=-\infty}^{\infty}e^{-i\omega_n\tau}\G(\k,i\omega_n)
 \end{eqnarray}
where $\nu_n=2n\pi/\beta$, $\omega_n=(2n+1)\pi/\beta$ with integer $n$ are the Matsubara frequencies.
Owing to the natural discretization of the Matsubara frequencies, it is more convenient to work with Matsubara frequencies.

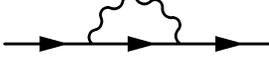
\begin{figure}
 \begin{fmffile}{diag1}
\begin{fmfgraph*}(100,50)
\fmfleft{i1}
\fmfright{o1}
\fmf{fermion,label=${\bf k}$,label.side=down}{i1,v1}
\fmf{fermion,label=${\bf k-q}$,label.side=down}{v1,v2}
\fmf{fermion,label=${\bf k}$,label.side=down}{v2,o1}
\fmf{photon,label=${\bf q}$,label.side=left,left=1.0,tension=0.003}{v1,v2}
\end{fmfgraph*}
\end{fmffile}
\caption[]{The first-order self-energy diagram.}
\label{diag-1}
\end{figure}

The matrix representation of the noninteracting Green's function in the Matsubara frequency and reciprocal space representation takes the following form for the electrons
  \begin{equation}
   \G_0(\k,i\omega_n) = [i\omega_n \mathbb{1} -\hat{\varepsilon}_{\k}\hat{S}]^{-1}
  \end{equation}
and for the phonons
  \begin{equation}
   D_{0\nu}({\bf q},i \omega_n)= \frac{-2\omega_{\q,\nu}}{\omega^2_{\q,\nu}+\omega^2_n}
  \end{equation}
The fully interacting electron and phonon single particle Green's
functions can be represented in terms of the non-interacting Green's function and the self-energy
stemming from the Dyson equation through
  \begin{eqnarray}\label{dyson}
   \G^{-1}({\k},i\omega_n) = \G^{-1}_0(\k,i\omega_n)-{\Sig}({\k},i\omega_n) \\
   {D_\nu}^{-1}({\q},i \nu_n) = {D_{0\nu}}^{-1}(\q,i \nu_n)- \Gamma_\nu(\q,i \nu_n)
  \end{eqnarray}
where ${\Sig}$ and $\Gamma$ are the electronic and the phononic self-energies.

Migdal's theorem \cite{migdal}
states that the vertex corrections to the electron self-energy are small,
hence setting the full vertex to the bare one is a reasonable approximation.
This particularly means that the interaction is truncated at order $\sqrt{m/M}\sim \omega_D/\epsilon_{\rm F}$, with $\omega_D$ is
Debye frequency, $\epsilon_{\rm F}$ is Fermi energy, $m$ and $M$ are bare electron and ionic masses, respectively.

The self-energy is approximated by the first-order diagram (see Fig.~\ref{diag-1}) in the Dyson series as the rainbow or non-crossing diagram.
The first order self-energy diagram given as follows

\begin{widetext}
 \begin{eqnarray}
  \Sig({\bf k},i \omega_n) = -T \int \frac{d{\bf k'}}{(2\pi)^2} \sum_{n'\nu} D_{0\nu}({\bf k-k'},i\omega_n-i\omega_{n'}) \hat{S} \hat{g}^\nu_{\bf k; k'} \hat{G}({\bf k'},i\omega_{n'}) \hat{S} \hat{g}^\nu_{\k'; \k}
 \end{eqnarray}
\end{widetext}
where $T$ is temperature and $d\k\equiv d^2k$. Apparently, the self-energy matrix is a $2N\times 2N$ matrix as well. Carefully looking at the structure of the Green's function,
one may represent it as a combination of the blocks of $2\times 2$ matrices which are
 labeled by combined band indexes $i$ and $j$. Notice that like $\hat{G}$ the self-energy matrix could be represented
 by a combination of blocks of $2\times 2$ matrices.
  Neglecting the inter-band contributions, the self-energy matrix will be block diagonal and could be represented by $2\times 2$ matrices which are labeled only by one band index, furthermore the $\hat{S}$ matrix is replaced by $\sigma_3$
\begin{widetext}
\begin{equation}\label{sg0}
  \Sig^{(1)}_{i}({\k}, i\omega_n) = -T \sum_{\nu jn'}\int \frac{d{\bf k'}}{(2\pi)^2} D_{0\nu}({\bf k-k'},i\omega_n-i\omega_{n'})
   |g^\nu_{{i\k};{j\k'}}|^2 \sigma_3\hat{G}_{j}({\bf k'},i\omega_{n'})\sigma_3
 \end{equation}
 \end{widetext}
with, $\sigma_3= \left( {\begin{array}{cc}1 & 0 \\  0 & -1 \\ \end{array} } \right)$,
 in this way the non-interacting Green's function reads,
 \begin{equation}\label{g022}
  \G_{0i}(\k,i\omega_n) = [i\omega_n \mathbb{1} -{\varepsilon}_{\k i}\sigma]^{-1}
 \end{equation}

\subsection{Isotropic approximation}

In order to simplify the problem, we adopt the averaging procedure which has been applied to the single band case, and we only focus on quantities near the Fermi surface.
To do so, we apply the averaging operator, i.e. $\displaystyle \frac{1}{N_i(0)}\int \frac{d{\bf k}}{(2\pi)^2}\delta(\varepsilon_{{\k i}})$ (the averaging operator is band dependent and $N_i(0)$ is the contribution of the band $i$ to the total DOS at the Fermi energy such that $N(0)=\sum N_i(0)$) on the both sides of the Eq.~(\ref{sg0}), which results in the following relation for the band-dependent self-energy,
  \begin{eqnarray}\label{sg0a}
  &&\Sig^{(1)}_i(i\omega_n) = \frac{T}{N_i(0)}\sum_{\nu jn'} \int \frac{d{\bf k} d{\bf k'}}{(2\pi)^4}\frac{2\omega_{{\bf k}-{\bf k}',\nu}}{(\omega_n-\omega_{n'})^2+{\omega_{{\bf k}-{\bf k}',\nu}}^2} \nonumber\\
  &&\times \delta(\varepsilon_{{\bf k} i}) {|g^{\nu}_{ {\bf k}i, {\bf k'}j}|}^2
    \sigma_3 \G_j({\bf k' },i\omega_{n'})\sigma_3
  \end{eqnarray}
where $\displaystyle \Sig_i(i\omega_n) = \langle\langle \Sig_i({\k },i\omega_n)\rangle\rangle_{FS}=\frac{1}{N_i(0)}\int \frac{d{\bf k}}{(2\pi)^2}\delta(\varepsilon_{\k i})\Sig_i({\bf k},i\omega_n)$.

Further simplification could be achieved by applying the averaging operator in integration over $\k'$ on the right-hand side of the
Eq.~(\ref{sg0a}), and further disentangling the integration over $\displaystyle g^{\nu}_{ {\bf k}i, {\bf k'}j}$ and
$\G_j({\bf k' },i\omega_{n'})$.
Having used those assumptions,
the k-independent self-energy now reads as,
  \begin{eqnarray}\label{sg0aa}
   &&\Sig^{(1)}_i(i\omega_n) = \sum_{\nu jn'} \frac{T}{N_i(0)N_j(0)} \int \frac{d{\bf k} d{\bf k'}}{(2\pi)^4} \frac{2\omega_{{\bf k}-{\bf k}',\nu}}{(\omega_n-\omega_{n'})^2+{\omega}^2_{{\bf k}-{\bf k}',\nu}} \nonumber \\
  &&\times{|g^{\nu}_{ {\bf k}i, {\bf k'}j}|}^2\delta(\varepsilon_{{\bf k} i}) \delta(\varepsilon_{{\bf k'} j}) \int\frac{d{\bf k'} }{(2\pi)^2}\sigma_3 \G_j({\bf k' },i\omega_{n'})\sigma_3 \nonumber \\
  \end{eqnarray}

This treatment relays on the fact that $\displaystyle g^{\nu}_{ {\bf k}i, {\bf k'}j}$ variations are smaller near
 the Fermi surface~\cite{allen2}. Therefore, it is replaced by its spherical average at the Fermi surface and large variations in the denominator of the
 $\G_j({\bf k' },i\omega_{n'})$ as a function of $\k'$ is treated exactly (see Eq.~(\ref{gloc})).
By rearranging Eq.~(\ref{sg0aa}), one may write down the following equation for the first-order self-energy
  \begin{eqnarray}\label{sg1}
   \Sig^{(1)}_i(i\omega_n) = T \sum_{jn'} \Lambda_{ij}(\omega_n-\omega_{n'}) \sigma_3 \G_j(i\omega_{n'})\sigma_3
  \end{eqnarray}
where $\Lambda_{ij}$ and $\G_j(\omega_{n'})$ are defined as
\begin{widetext}
  \begin{eqnarray}
&&\G_j(i\omega_{n}) =  \int\frac{d\k }{(2\pi)^2}  \G_j(\k,i\omega_{n})= \int \frac{d\k }{(2\pi)^2} [i\omega_n \mathbb{1} -\varepsilon_{\k j}\sigma_3-\Sig_j(i\omega_n)]^{-1} = \int d\varepsilon N_j(\varepsilon)[i\omega_n \mathbb{1} -\varepsilon\sigma_3-\Sig_j(i\omega_n)]^{-1} \label{gloc}\\
&&\Lambda_{ij}(\omega_n-\omega_{n'}) = \frac{1}{N_j(0)} \int d\Omega \frac{2\Omega\alpha^2\F_{ij}(\Omega)}{(\omega_n-\omega_{n'})^2+{\Omega}^2}  \label{lamomega}\\
 &&\alpha^2\F_{ij}(\Omega) = \frac{1}{N_i(0)}\sum_{\nu}\int\frac{d{\bf k}}{(2\pi)^2}\frac{d{\bf k'} }{(2\pi)^2} {|g^{\nu}_{ {\bf k}i, {\bf k'}j}|}^2 \delta(\varepsilon_{{\bf k} i})\delta(\varepsilon_{{\bf k'} j})\delta(\Omega-\omega_{{\bf k}-{\bf k}',\nu})
 \label{a2fij0}
 \end{eqnarray}
 \end{widetext}
while $\alpha^2\F_{ij}$ is not symmetric within the exchange of the indexes, $\Lambda_{ij}$ is a symmetric function within the exchange of $i$ and $j$ indexes.
Considering only the first-order diagram, i.e. $\Sig_i(i\omega_n)=\Sig^{(1)}_i(i\omega_n)$, the self-consistent solution of Eq.~(\ref{sg1}) with the Dyson equation (Eq.~(\ref{dyson})) completes the solution of the Eliashberg equations. Furthermore,
it is a common practice to parameterize the momentum averaged self-energy as,
\begin{eqnarray}\label{selfmultip}
    \Sig_i(i\omega_n) &=& i\omega_n[1-Z_i(\omega_{n})]\mathbb{1} + \chi_i(\omega_{n})\sigma_3  \nonumber \\
                        &+& \phi_i(\omega_{n})\sigma_1 + \bar{\phi_i}(\omega_{n})\sigma_2
                        \end{eqnarray}
where $\sigma_1= \left( {\begin{array}{cc}0 & 1 \\  1 & 0 \\ \end{array} } \right)$, $\sigma_2= \left( {\begin{array}{cc}0 & -i \\   i & 0 \\ \end{array} } \right)$ and $\mathbb{1}$ is a $2\times 2$ unit matrix.
In the following, we choose a gauge such that $\bar{\phi}_i(\omega_{n})=0$~\cite{allen2}. By inserting the above mentioned
decomposition for the local self-energy into the Dyson equation Eq.~(\ref{dyson}) and taking into account that  $\bar{\phi_i}(\omega_{n})=0$,
$\G_i({\bf k },\omega_{n})$ reads
\begin{eqnarray}
  \G_i({\bf k},i\omega_n) =&-&[i\omega_nZ_i(i\omega_{n}) \mathbb{1} + (\epsilon_{\bf k}+\chi_i(i\omega_{n}))\sigma_3 \nonumber \\
                          &+&\phi_i(i\omega_{n})\sigma_1]/{\Theta_i(i\omega_{n})}
\end{eqnarray}
with $\Theta_i(i\omega_n)= (\omega_nZ_i(i\omega_n))^2 + (\epsilon_{\bf k}+\chi_i(i\omega_n))^2+\phi_i^2(i\omega_n)$.
By using the mentioned parametrization of $\G_i({\bf k},i\omega_n)$, considering the definition of $\G_i(i\omega_{n})$ and implying the DOS, which is a constant for the bands at the Fermi energy,
 one may calculate the integral over $\varepsilon$ (in Eq. (\ref{gloc})) analytically.
 Performing the integration, $\G_i(i\omega_{n})$ becomes
\begin{eqnarray}
 \G_i(i\omega_{n}) &=&-\pi N_i(0) \frac{[i\omega_nZ_i(i\omega_n) \mathbb{1}  +\phi_i(i\omega_n)\sigma_1]}{\Xi_i(i\omega_n)}\nonumber\\
  \Xi_i(i\omega_n)&=&\sqrt{  [\omega_n Z_i(i\omega_n)]^2 + [\phi_i(i\omega_{n})]^2}
\end{eqnarray}

Inserting the above equation back to Eq.~(\ref{sg1}) and equating the corresponding elements of the matrices on both sides,
the following set of equations for the components of the self-energy is achieved~\cite{nicol},

\begin{eqnarray}
&&Z_i(i\omega_n)  = 1+ \frac{\pi T}{\omega_n}\sum_{jn'} \lambda_{ij}(\omega_n-\omega_{n'})\frac{\omega_{n'}Z_{j}(i\omega_{n'})}{\Xi_{j}(i\omega_{n'})}\nonumber\\
&&\phi_i(\omega_n) = \pi T\sum_{jn'} \lambda_{ij}(\omega_n-\omega_{n'})\frac{\phi_{j}(i\omega_{n'})}{\Xi_{j}(i\omega_{n'})}
\end{eqnarray}
with $\lambda_{ij}(\omega_n-\omega_{n'})=N_{j}(0)\Lambda_{ij}(\omega_n-\omega_{n'})$.

\subsection{Projected quantities}\label{projections}

For the illustrative purposes, we compute the different projections of quantities like $\alpha^2\F$ and ${\bf F}$. We define the projected phonon DOS ${\bf F}_\kappa(\omega)$ into the
Cartesian coordinates $\kappa$, which represents the contribution of the phonons with polarization in the $\kappa$ direction to the total phononic DOS. We consider only two major directions, an in-plane ($\overline{xy}$) and out-of-plane ($\overline{z}$),
\begin{equation}
 {\bf F}_\kappa(\omega) = \sum_{\nu}\int \frac{d{\bf q}}{(2\pi)^2} {\bf P}^{\nu\q}_\kappa\delta(\omega-\omega_{\q,\nu})
\end{equation}
where $\kappa=\overline{xy}$ and $\overline{z}$, ${\bf P}^{\nu\q}_{\overline{xy}}={\displaystyle\sum_s\sum_{\kappa=x,y} } {{\bf e^{\q\nu}}_{s\kappa}}{{{\bf e^{\q\nu^*}}_{s\kappa}}}$ and
${\bf P}^{\nu\q}_{\overline{z}}={\displaystyle \sum_{ s} } {\bf e}^{\q\nu}_{s \overline{z}}{{\bf e}^{\q\nu^*}_{s \overline{z}}}$, $s$ is
 the atomic index in the unit-cell, and ${\bf e}^{\q\nu}$ is phonon polarization for branch $\nu$ and
 vector $\q$ (see Appendix~\ref{sec:appa} for the definition), and ${\bf P}^{\nu\q}_\kappa$ satisfies $\sum_{\kappa=\overline{xy},\overline{z}}{\bf P}^{\nu\q}_\kappa=1$. 

The second projected quantity is projected $\alpha^2\F$ into the Cartesian directions of the phonon displacements.
The quantity is used to identify the contribution of the phonons with a specific character (here Cartesian
displacement of phonons) in $\alpha^2\F$ and therefore $\lambda$,
\begin{eqnarray}
   \alpha^2{ \bf F}^{\kappa\kappa'}_{ij}(\Omega) = \frac{1}{N_i(0)}\int\frac{d{\bf k}}{(2\pi)^2}\frac{d{\bf k'} }{(2\pi)^2} {g^{\nu\kappa}_{ {\bf k}i, {\bf k'}j}}
   {g^{\nu\kappa'^*}_{ {\bf k}i, {\bf k'}j}}\nonumber \\
  \times \delta(\varepsilon_{{\bf k} i})\delta(\varepsilon_{{\bf k'} j})\delta(\Omega-\omega_{{\bf k-k'},\nu})
  \end{eqnarray}
$\kappa,\kappa'=\overline{xy},\overline{z}$ and ${\displaystyle g^ {\nu \overline{xy}}_{ {\bf k}i, {\bf k'}j}}= {\displaystyle\sum_{\kappa=x,y}g^{\nu\kappa}_{ {\bf k}i, {\bf k'}j}}$
(for the definition of $g^{\nu\kappa}_{ {\bf k}i, {\bf k'}j}$ see Appendix~\ref{sec:appa}).
Hence, we have the following relations between the projected $\alpha^2\F$s as
 \begin{eqnarray}\label{a2fprojs}
 &&\alpha^2\F_{ij}(\Omega) = {\displaystyle \sum_{\kappa,\kappa'=\overline{xy},\overline{z}} \alpha^2\F^{\kappa\kappa'}_{ij}(\Omega)} \label{a2fij}\\
  &&\displaystyle \alpha^2\F^{\kappa\kappa'}(\Omega)=\frac{1}{N(0)} \sum_{ij} N_i(0)\alpha^2 \F^{\kappa\kappa'}_{ij}(\Omega) \label{a2fkk}\\
  \end{eqnarray}
  The total $\alpha^2{\bf F}$ reads
  \begin{eqnarray}
  &&\displaystyle \alpha^2\F(\Omega)=\frac{1}{N(0)} \sum_{ij} N_i(0)\alpha^2 \F_{ij}(\Omega)\label{a2ftot}
 \end{eqnarray}

The band projected mass renormalization factor is defined as
\begin{equation}
 \lambda_{ij}= 2\int d\Omega \frac{\alpha^2\F_{ij}(\Omega)}{\Omega}
\end{equation}
 the relation between total $\lambda$ and $\lambda_{ij}$ is $\lambda = \frac{1}{N(0)}\sum_{ij} N_i(0)\lambda_{ij}$.
\begin{figure}
\begin{fmffile}{diag2}
\begin{fmfgraph*}(200,90)
\fmfleft{i1}
\fmfright{o1}
\fmf{fermion,label=\tiny$\bf k$,label.side=down}{i1,v1}
\fmf{fermion,label=\tiny$\bf k-q$,label.side=down}{v1,v2}
\fmf{fermion,label=\tiny$\bf k-q-q'$,label.side=down}{v2,v3}
\fmf{fermion,label=\tiny$\bf k-q'$,label.side=down}{v3,v4}
\fmf{fermion,label=\tiny$\bf k$,label.side=down}{v4,o1}
\fmf{photon,label=\tiny$\bf q'$,label.side=up,left=1.0,tension=0.003}{v2,v4}
\fmf{photon,label=\tiny$\bf q$,label.side=up,left=1.0,tension=0.003}{v1,v3} 
\end{fmfgraph*}
\end{fmffile}
\caption[]{Correction  of  the second  order  self-energy in the  electron-
phonon interaction to the electron propagator.}
\label{diag-2}
\end{figure}
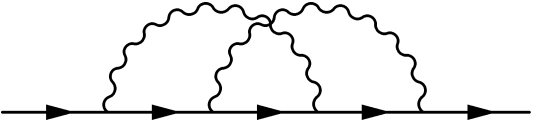

\subsection{Coulomb Interaction contribution}
In the following, we only focus on the band with the largest contribution on
 $\alpha^2{\bf F}$ (we will show that the main contribution comes from band 5, where band 5 is marked with green color in Fig.~\ref{dosband}).
The justification of this assumption is based on the fact that the contribution of band 4 in $\alpha^2{\bf F}$ is negligible in comparison to that from band 5 for the largest Fermi energy shift (see Fig.~\ref{a2fprojbnd} in numerical section) and the other lower bands
acquire vanishing $\alpha^2{\bf F}$ for all examined $\delta E_{\rm F}$ (not shown here). Therefore,
we only consider band 5 and for the sake of simplicity, we neglect band indexing regarding band structure, Green's function and self-energies.

The Hamiltonian corresponding to the electron-electron interaction has the following form~\cite{ummarino,scalapino},
   \begin{widetext}
   \begin{equation}
    \hat{\cal{H}}_c = \frac{1}{2}\sum_{{\bf k}_1 {\bf k}_2 {\bf k}_3 {\bf k}_4}
    \langle {\bf k}_3 {\bf k}_4|V_{C}|{\bf k}_1 {\bf k}_2\rangle \psi^\dagger_{{\bf k}_3}\sigma_3\psi_{ {\bf k}_1}\times\psi^\dagger_{{\bf k}_4}\sigma_3\psi_{{\bf k}_2}
   \end{equation}
   \end{widetext}
$\psi_{\k}= \left( {\begin{array}{cc} c_{\k\uparrow} \\  c^\dagger_{-\k\downarrow}\\ \end{array} } \right)$, $c_{\k\sigma}$ ($c^\dagger_{\k\sigma}$) annihilates (creates) electrons
 in band 5, reciprocal vector $\k$ and spin $\sigma$. $\langle {\bf k}_3 {\bf k}_4|V_{C}|{\bf k}_1 {\bf k}_2\rangle$ is the bare electron-electron Coulomb interaction, the translational
 invariance of $V_c$ restricts ${\bf k}_1+{\bf k}_2-{\bf k}_3- {\bf k}_4$ being zero or a reciprocal lattice vector ${\bf K}$. Considering the above Coulomb contribution to $\hat{\cal{H}}_0$ and neglecting
  the contribution of the other bands, one may write the isotropic first order self-energy as follows,
   \begin{eqnarray}\label{s1}
   \Sig^{(1)}(i\omega_n) = T \sum_{n'} \{ \Lambda(\omega_n-\omega_{n'}) -\bar{V}_C \}\sigma_3 \G^{od}(i\omega_{n'})\sigma_3 \nonumber \\
  \end{eqnarray}
with $\displaystyle  \bar{V}_C = \frac{1}{{N(0)}^2}\int \frac{d{\bf k}}{(2\pi)^2}\frac{d{\bf k'} }{(2\pi)^2} V_C({\bf k-k'})\delta(\varepsilon_{{\bf k'} })\delta(\varepsilon_{{\bf k} })$,
$\displaystyle \Lambda(\omega_n-\omega_{n'}) = \frac{1}{N_5(0)} \int d\Omega \frac{2\Omega\alpha^2\F_{55}(\Omega)}{(\omega_n-\omega_{n'})^2+{\Omega}^2}$,
and $\G^{od}$ holds for off-diagonal part of the Green's function. The reason of holding only the off-diagonal part of Green's function is the fact that the electron-electron interaction has already been considered for diagonal parts in DFT calculation of the electronic band structure~\cite{allen2}.
By imposing the constant DOS approximation, the isotropic Eliashberg equations have the following form
\begin{eqnarray} \label{eliashbergsingle}
&&\omega_n[1-Z(i\omega_n)]  = -\pi T\sum_{n'} N(0)\Lambda(\omega_n-\omega_{n'})\frac{\omega_{n'}Z(i\omega_{n'})}{\Xi(i\omega_{n'})} \nonumber\\
&&\phi(i\omega_n)=\pi T\sum_{n'} [N(0) \Lambda(\omega_n-\omega_{n'})
                - N(0)\bar{V}_c ] \frac{\phi(i\omega_{n'})}{\Xi(i\omega_{n'})}\nonumber \\
\end{eqnarray}

In spite of the electron-phonon interaction kernel, the $\bar{V}_C$ does not have any natural upper cutoff in energy summation in Eq.~(\ref{eliashbergsingle}).
However, owing to retardation effects, the repulsion felt by the electrons is smaller than instantaneous interactions. The procedure of the scaling of the Coulomb interaction is to replace the $\mu_c=N(0)\bar{V}_C$ with the well-known Morel-Anderson pseudopotential $\displaystyle \mu^*_c = \frac{\mu_c}{1+\mu_c \ln(E/\omega_D)}$
with $E$ is the electronic bandwidth and $\omega_D$ is the phonon
energy scale which is as order of Debye energy~\cite{sch,bogo,morel,marsgilio2}. These retardation effects are still operative
for larger interactions and where higher order corrections are necessary, although they are less efficient due
to the reduction in bandwidth~\cite{bauer}. Furthermore, retardation effects impose an upper energy cutoff for
$\mu^*_c$ in energy summation~\cite{marsgilio2} in Eq.~(\ref{eliashbergsingle}). Eventually, by considering the above mentioned
retardation effects of electron-electron interaction, one may rewrite Eq.~(\ref{eliashbergsingle}) as

   \begin{eqnarray} \label{eliashbergsingle-muc}
&&Z(i\omega_n)  = 1+\frac{\pi T}{\omega_n}\sum_{n'} \lambda(\omega_n-\omega_{n'})\frac{\omega_{n'}Z(i\omega_{n'})}{\Xi(i\omega_{n'})} \nonumber\\
&&\phi(i\omega_n)=\pi T\times \sum_{n'} [\lambda(\omega_n-\omega_{n'}) - \mu^*_c \varTheta(\omega_c-|w_{n'}|)] \frac{\phi(i\omega_{n'})}{\Xi(i\omega_{n'})} \nonumber \\
\end{eqnarray}
with $\lambda(\omega_n-\omega_{n'})=N(0)\Lambda(\omega_n-\omega_{n'})$,
$\displaystyle \varTheta(x)$ is the Heaviside step function and $\omega_c \sim 5-10~ \omega_D$.
In order to achieve a common ground for comparison between the constant DOS approximation and that of
 variable DOS at the Fermi energy, in analogy with the constant DOS approximation, one may express Eq.~(\ref{s1}) as
\begin{eqnarray}\label{s1-od-muc}
   \Sig^{(1)}(i\omega_n) &=& T \sum_{n'} \{ \Lambda(\omega_n-\omega_{n'}) \sigma_3 \G(i\omega_{n'})\sigma_3 \nonumber \\
   &-& \frac{\mu^*_c}{N(0)}\varTheta(|w_{n'}|-\omega_c) \sigma_3 \G^{od}(i\omega_{n'})\sigma_3 \}
  \end{eqnarray}

Clearly, by using constant DOS approximation for Eq.~(\ref{s1-od-muc}) together with Dyson equation, one finds
Eqs.~(\ref{eliashbergsingle-muc}).

\subsection{Second-order self-energy: Vertex corrections}
When there is a flat band near the VBM
(as is the case for us corresponding to the electronic structure of the BLP, see Fig.~\ref{dosband}),
the first-order self-energy might not be accurate enough to explore the physics of the system.
In this case, we should include the second-order self-energy considering the vertex corrections~\cite{cappelluti}.
In addition, since we incorporate all electron-electron contributions in the electron Green's function, therefore,
we will just consider the vertex correction on the electron-phonon interaction.
To proceed, we impose further simplifications
to overcome difficulties owing to the computational complexity of the problem. Therefore, by using the calculated $\alpha^2\F$ for each Fermi energy shift, we derive an effective Hamiltonian
such that the interaction kernel of the isotropic averaged interaction kernel is the same as the interaction kernel of the newly constructed Hamiltonian. In this regard, we consider a simple Holstein
model composed of the two dispersionless Einstein modes. Furthermore, we consider a model which has the same electronic band structure as the band structure of band 5,

 \begin{widetext}
 \begin{eqnarray}
  \hat{\cal{H}} = \sum_k \varepsilon_\k \psi^\dagger_\k\sigma_3\psi_\k + \sum_{\nu=1,2}\sum_{\q} \omega_{\nu}b^\dagger_{\q,\nu}b_{\q,\nu} +\sum_{\nu=1,2} \sum_{\k,\q} g_{\nu}(b^\dagger_{\q,\nu}+b_{-\q,\nu}) \psi^\dagger_{\k+\q}\sigma_3\psi_{\k} \nonumber\\
           +\frac{1}{2}\sum_{\k_1 \k_2 \k_3 \k_4} \langle \k_3 \k_4|V_{C}|\k_1 \k_2\rangle \psi^\dagger_{\k_3}\sigma_3\psi_{ \k_1}\times\psi^\dagger_{\k_4}\sigma_3\psi_{\k_2}
 \end{eqnarray}
 \end{widetext}
where $\varepsilon_{\k}$ is the energy dispersion of band 5 relative to $E_{\rm F}$, $\omega_\nu$ is the phonon energy and $g_\nu$ is the corresponding electron-phonon interaction. The interaction kernel of this model can be written as $\tilde{\Lambda}(\omega_n-\omega_{n'})=\sum_\nu\frac{2\omega_{\nu}{g_\nu}^2}{{(\omega_n-\omega_{n'})}^2+{\omega_{\nu}}^2}$. In the next step, we fit the kernel of the new system
with that of the original system. Happily, we find that the kernel of the
system with two phonon modes fits with great accuracy to the kernel of the original model.
In order to include vertex corrections, therefore, we evaluate the second crossing self-energy diagram (see Fig.~\ref{diag-2}) like as,
\begin{widetext}
\begin{eqnarray}
 \Sig^{(2)}({\bf k},i\omega_n) = {(-T)}^2 \sum_{\omega_{n_1},\omega_{n_2}}\int \frac{d{\bf q} }{(2\pi)^2}\frac{d{\bf q'} }{(2\pi)^2} \sigma_3 \G({\bf k-q},i\omega_{n1})\sigma_3\G({\bf k-q-q'},i\omega_{n2})\nonumber\\
 \times\sigma_3\G({\bf k-q'},i\omega_{n}-i\omega_{n1}+i\omega_{n2})
  \sigma_3\Lambda(\omega_{n}-\omega_{n1})\Lambda(\omega_{n1}-\omega_{n2})
\end{eqnarray}

By averaging over the Fermi surface, the self energy is given by
\begin{eqnarray}\label{ss21a}
 \Sig^{(2)}(i\omega_n) = \frac{{(-T)}^2}{N(0)} \sum_{\omega_{n_1},\omega_{n_2}}\int \frac{d{\bf k} }{(2\pi)^2} \frac{d{\bf q} }{(2\pi)^2}\frac{d{\bf q'} }{(2\pi)^2} \delta(\varepsilon_{\bf k})\sigma_3
  \G({\bf k-q},i\omega_{n1})\sigma_3  \G({\bf k-q-q'},i\omega_{n2})\nonumber \\ \times\sigma_3\G({\bf k-q'},i\omega_{n}-i\omega_{n1}+i\omega_{n2})\sigma_3
  \Lambda(\omega_{n}-\omega_{n1})\Lambda(\omega_{n1}-\omega_{n2})
\end{eqnarray}
\end{widetext}

The evaluation of the Eq.~(\ref{ss21a}) has a computational complexity as order of $O(N^6)$ regarding reciprocal integration, where $N$ is the number of mesh points in each direction of the reciprocal space. To reduce the complexity,
it is more efficient to evaluate the diagram in the real space. To perform the evaluation of the Eq.~(\ref{ss21a}), we use the real space representation of the second order-diagram followed by a backward Fourier transformation on the real-space diagram. The real-space diagram can be evaluated as follows
 \begin{eqnarray}\label{ss21r}
 &&\Sig^{(2)}({\bf x},i\omega_n) = {(-T)}^2 \sum_{\omega_{n_1},\omega_{n_2}} \sigma_3 \G({\bf x},i\omega_{n_1})\sigma_3\G({\bf -x},i\omega_{n_2})
 \sigma_3 \nonumber \\
 &&\times \G({\bf x},i\omega_{n}-i\omega_{n_1}+i\omega_{n_2}) \sigma_3
\Lambda(\omega_{n}-\omega_{n1})\Lambda(\omega_{n1}-\omega_{n2})
 \end{eqnarray}
 where $\G({\bf x},i\omega_{n})=\int \frac{d{\bf k} }{(2\pi)^2} e^{-\imath {\bf x}\cdot{\bf k}}\G({\bf k},i\omega_{n})$ and the second-order self-energy is
\begin{equation}
 \Sig^{(2)}(i\omega_n) = \frac{1}{N(0)}\int \frac{d{\bf k} }{(2\pi)^2}\delta(\varepsilon_{{\bf k }})\Sig^{(2)}({\bf k},i\omega_n)
\end{equation}
with $\displaystyle \Sig^{(2)}({\bf k},i\omega_n)=\sum_x e^{i\k . x}\Sig^{(2)}({\bf x},i\omega_n)$. Having calculated the first-and second-order self-energy, the total self-energy is thus given by
\begin{equation}\label{ss21}
 \Sig(i\omega_n) = \Sig^{(1)}(i\omega_n)  + \Sig^{(2)}(i\omega_n)
\end{equation}
As usual, a self-consistency must be imposed between Eq.~(\ref{ss21}) and Dyson equation, Eq. (\ref{dyson}).

\section{Numerical Results}\label{sec:result}

In this section we mainly consider the Migdal-Eliashberg formalism considering the first-order self-energy.
Eventually, we also include the vertex correction to the self-energy and only discuss the superconducting energy
gap and critical temperature as well. We define $\varepsilon_{\bf k}=\tilde{\varepsilon}_{\bf k}- E_{\rm F}$,
where $\tilde{\varepsilon}_{\bf k} $ is the energy dispersion, which is extracted from DFT calculations. Once again,
we investigate BLP under the rigid shift of the Fermi energy deep into the lower energies below the VBM.
In particular, our analysis is based on $\delta E_{\rm F} = -0.02,-0.055,-0.105, -0.155$ and $-0.205$ eV.
The corresponding hole densities are $\rho_h=5.0\times10^{13}$, $1.5\times10^{14} $, $2.4\times10^{14}$ , $3.1\times 10^{14} $ and $3.8\times10^{14}$ cm$^{-2}$, respectively,
with $\displaystyle \rho_h(E_{\rm F})=[N_{tot}-\int_\infty^{E_{\rm F}}N(\varepsilon)d\varepsilon]/S$, where $N(\varepsilon)$ is the DOS, $S$ is the unit-cell surface area and $N_{tot}$ is the
 total holes in the valence band when $E_{\rm F}=E_{VBM}$.

The interpolation is performed for a range of fine meshes. In most critical cases, we use a fine $800\times 800$ $k$-mesh and $200 \times 200$ $q$-mesh,
whereas in less critical cases, we use $200\times 200$ $k$-mesh and $200 \times 200$ $q$-mesh. The delta functions are approximated by a Gaussian function as
$\displaystyle \delta(x)\sim\frac{1}{\sqrt{\pi}\sigma}e^{-x^2/{\sigma}^2}$ for calculating the $\alpha^2{\bf F}$ and $\lambda$. The convergence of the quantities
is thoroughly checked for a range of $\sigma$, $k$- and $q$-meshes (see Appendix~\ref{sec:appb}). Particularly, the DOS, $\alpha^2{\bf F}$ and $\lambda$ are insensitive
to the electronic broadening for a range $\sigma=0.00125-0.01$ eV and the phononic broadening is set to be $\sigma_{ph}=0.1$ meV. Furthermore, due to the small energy scale of the out-of-plane acoustic mode, inaccuracies are inevitable. In order to filter numerical inaccuracies out, for actual calculations of $\lambda$ and solutions of the Eliashberg equations a lower cutoff $\sim 1$ MeV was considered such that below this cutoff the $\alpha^2{\bf F}$ is omitted.
  To further reduce the numerical complexity of the evaluation of Eq.~(\ref{ss21r}), we employ the observation that Green's function is a quiet local in real space, therefore, an upper real-space cutoff over which the
 $\Sig^{(2)}({\bf x},i\omega_n)$ was set to zero, is considered. Here, we use a cutoff of 20 sites for each direction. Moreover, while for the evaluation of the Eq.~(\ref{ss21r}) we use $\alpha^2{\bf F}$ resulting from finest mesh available, all the vertex corrections are performed on a $200 \times 200$ $k-$mesh with $\sigma=0.01$ EV and the upper cutoff
  in frequency summation $\omega_c=0.5$ eV.

\begin{figure}[h]%
   \includegraphics*[width=0.4\textwidth]{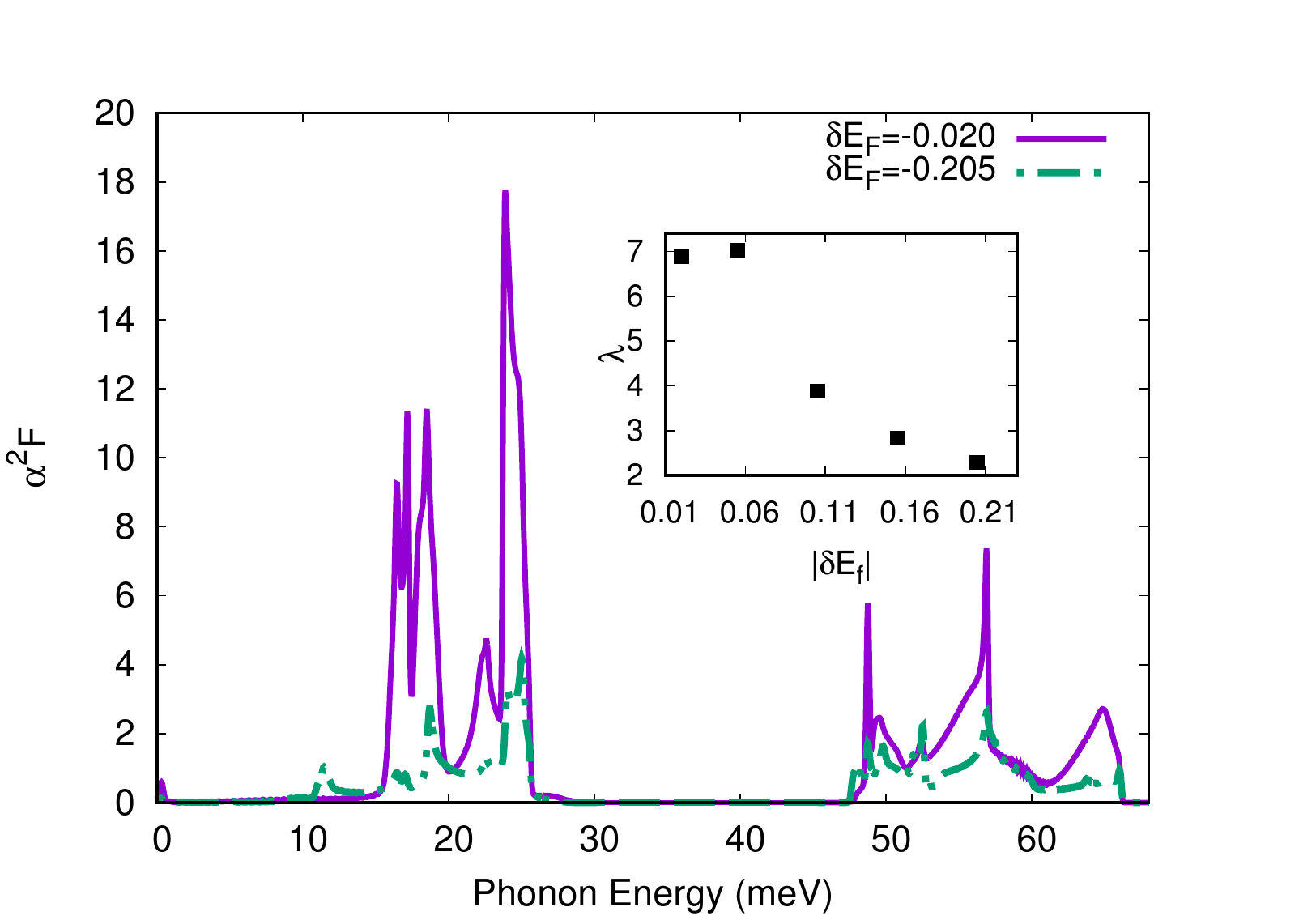}%
  \caption[]{%
     (Color online) Total $\alpha^2 \F$ as a function of the rigid shift of the Fermi energy for different values of
     $\delta{E_f}=-0.02$ and $-0.205$ eV. Notice that in the case of the large shifts,
     total $\alpha^2 \F$ suffers from a dramatic reduction in its magnitude since a reduction in the DOS induces a decreasing in the
     $\lambda$. Inset: $\lambda$ as a function of the rigid shift of the Fermi energy for different values of $\delta E_{\rm F}$.}
    \label{a2fvsef}
\end{figure}

Fig.~\ref{a2fvsef} depicts the total $\alpha^2 \F$ for different rigid shift of the Fermi energy, $\delta E_{\rm F} = -0.02$ and $-0.205$ eV, deep into the valence states.
Apparently for the larger shift, $\alpha^2 \F$ suffers from a dramatic reduction in its magnitude.

As it is obvious from the inset of Fig.~\ref{a2fvsef}, the total unit-less coupling, $\lambda$, shows a dramatic decreasing as a
function of $\delta{E_{\rm F}}$. Looking at the form of the $\alpha^2\F$ (see Eq.~(\ref{a2ftot})), it is expected that by reducing the DOS, the $\lambda$ decreases as well.
This could be qualitatively attributed by considering a dispersion-less phonon spectrum. $\lambda$ can be evaluated as $\lambda= 2N(0)\tilde{g}^2/\omega_0$, where $\omega_0$ and $\tilde{g}$ are the effective dispersion-less phonon energy and electron-phonon interaction for each Fermi energy shift and $N(0)$ is the DOS at the Fermi energy. Therefore, a reduction in the DOS induces a decreasing in the $\lambda$ as long as $\tilde{g}$ is a constant as a function of the $\delta E_{\rm F}$.

\begin{figure}
  \includegraphics*[width=0.47\textwidth]{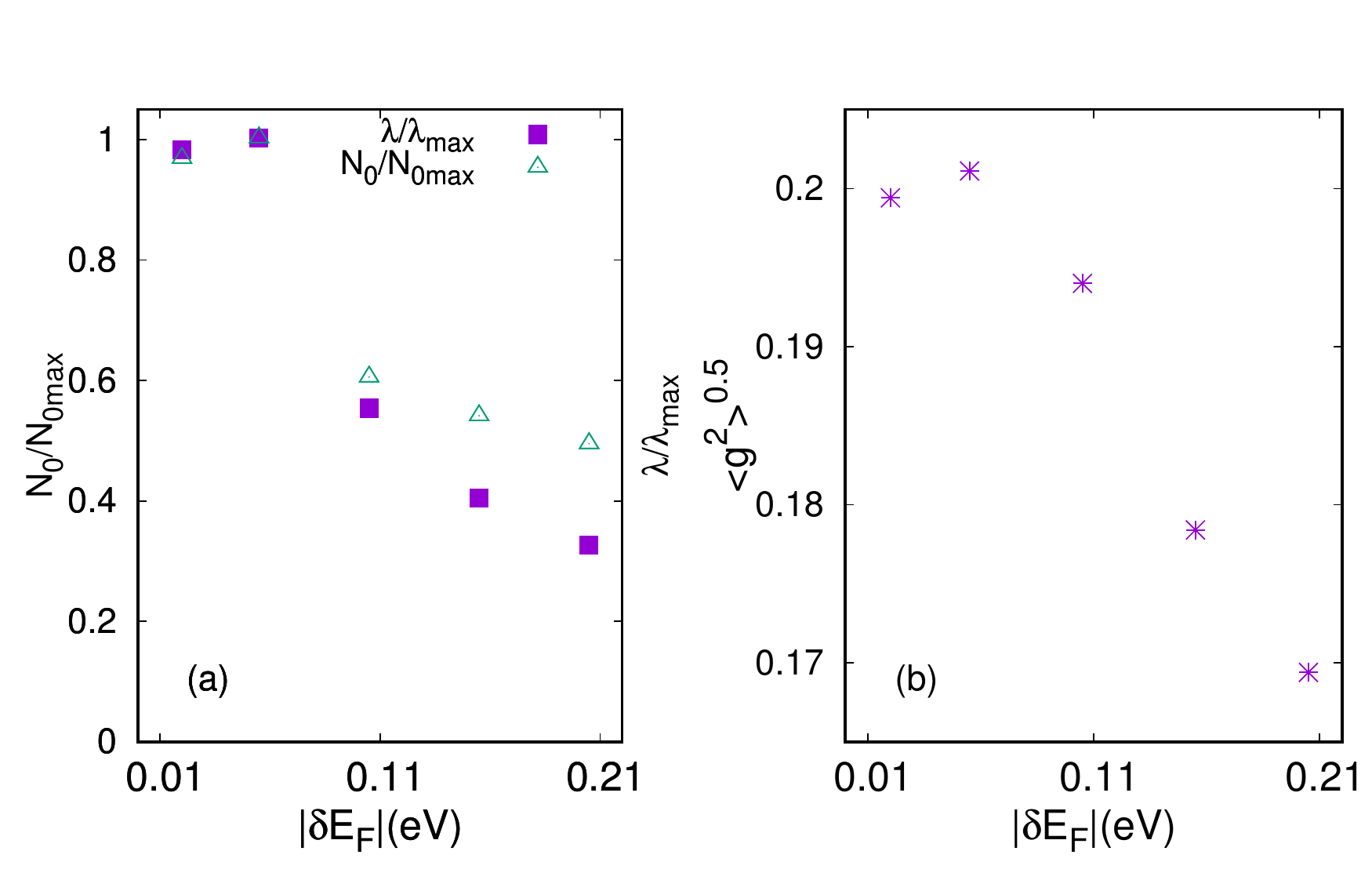}%
  \caption[]{%
      (Color online) (a) $\lambda/\lambda_{\rm max}$, $N(0)/N(0)_{\rm max}$ and (b) $\sqrt{\langle g^2\rangle}$ as a function of $\delta{E_{\rm F}}=-0.02,-0.055,-0.105,-0.155$ and $-0.205$ eV.
      Upon larger $\delta E_{\rm F}$ both $\lambda_{\rm max}$ and $N(0)$ decrease but the faster decrease
       is obvious for $\lambda$, which is the signature of the decreasing in averaged electron-phonon interaction (see part (b) of the figure).}
     \label{lamgavgn0}
\end{figure}

In order to disentangle the share of the $N(0)$ and $\tilde{g}$ for different $\lambda$ as a function of $\delta E_{\rm F}$,
in Fig.~\ref{lamgavgn0}(a), we plot $\lambda/\lambda_{\rm max}$ and $N(0)/N(0)_{\rm max}$ as a function of $\delta E_{\rm F}$.
By moving into the valence band states, both $\lambda$ and $N(0)$ decrease as a function of the $\delta E_{\rm F}$, however, the rate of the decreasing in $\lambda$ is larger than that of the $N(0)$ when they are compared with the $\lambda_{\rm max}$ and $N(0)_{\rm max}$, where generally it is the signature of the suppression of the electron-phonon interactions. Therefore, we plot $\sqrt{\langle g^2\rangle}$ as a function of the $\delta E_{\rm F}$ in Fig.~\ref{lamgavgn0}(b), where $\displaystyle \langle g^2\rangle=\frac{1}{N(0)}\int \alpha^2\F(\Omega)d\Omega$. By moving into the valence band states, the value of the $\sqrt{\langle g^2\rangle}$ suppresses. Hence, the reduction in the unit-less electron-phonon coupling as a function of the $\delta E_{\rm F} $ is not only owing to the suppression of the DOS at the Fermi energy, but also the electron-phonon interaction is generally suppressed when one changes $E_{\rm F} $ deeper into the valence bands.

In order to observe which character of the system is responsible for a behavior of the $\lambda$ as a function of the $\delta E_{\rm F}$,
we look at the projected $\alpha^2 \F$ into Cartesian directions of the phonon
displacements. The quantities are already defined in Sec.~\ref{projections}. We consider two major projection directions,
i.e. $\kappa,\kappa' = \overline{xy},\overline{z}$.

\begin{figure}[h]%
  \includegraphics*[width=0.4\textwidth]{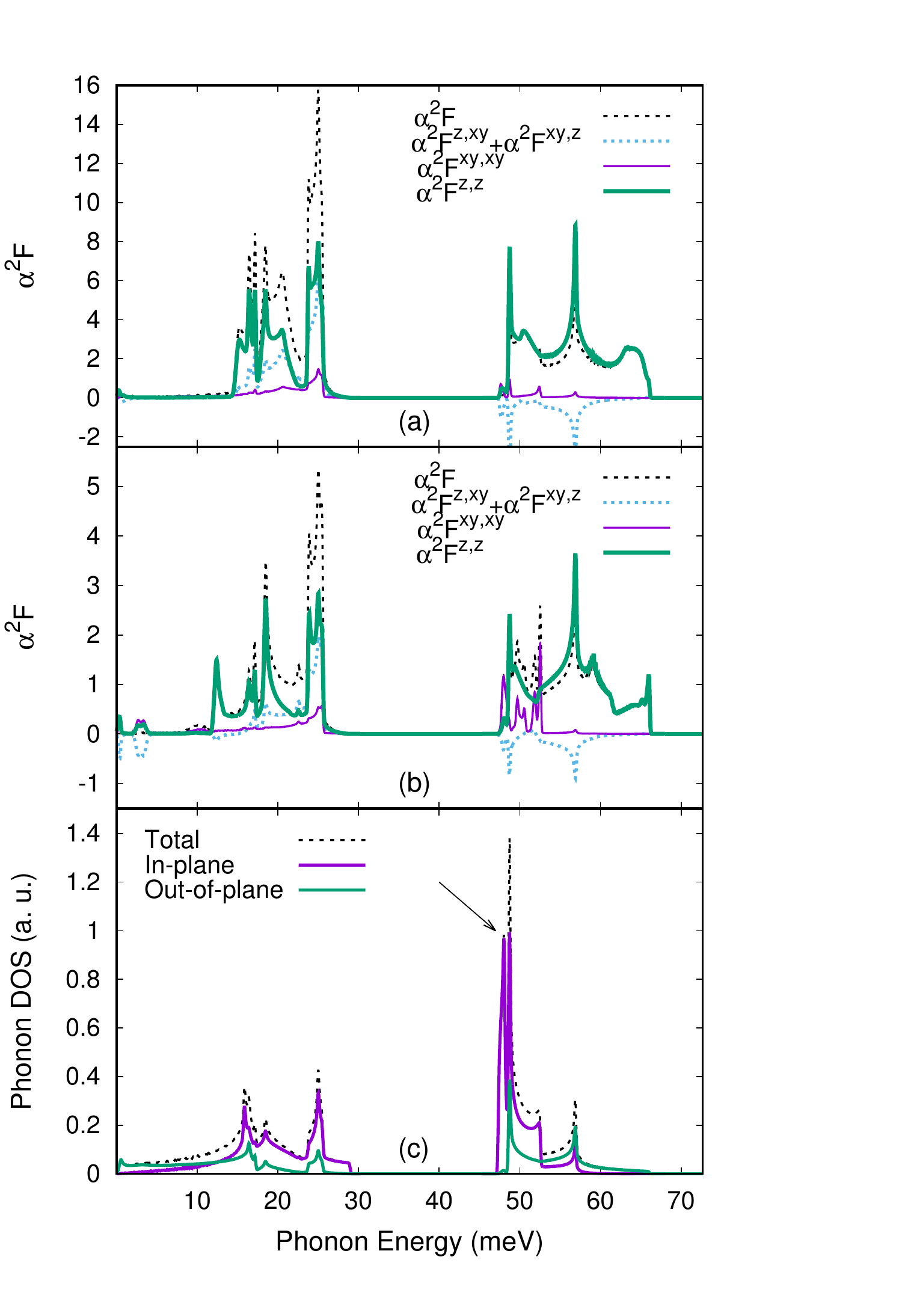}
  \caption[]{%
    (Color online) (a) and (b) Projected $\alpha^2\F$ for $\delta{E_{\rm F}}=-0.055$ and $\delta{E_{\rm F}}=-0.205$ eV respectively, (c) Projected phonon DOS.
     The electrons couple to the deformations which are induced
 the out-of-plane displacements of the phonons. In the presence of the phonons with considerable out-of-plane
character is a quantity to acquire sizable $\alpha^2 \F$.}
    \label{a2f07}
\end{figure}

In Fig.~\ref{a2f07}(a) the projected $\alpha^2 \F^{\kappa\kappa'}$ as a function of phonon energy
for $\delta E_{\rm F}=-0.055$ eV is presented for different projection directions.
It is seen that the electrons mainly couple to the deformations which are induced
by the out-of-plane displacements of the phonons. Comparing with the projected phonon DOS in Fig.~\ref{a2f07}(c),
it is possible to observe that the presence of the phonons with considerable out-of-plane
character is a quantity to acquire sizable $\alpha^2 \F$. In particular, by noticing at the lower edge of the optical
 phonon spectrum, around 48 meV which is indicated by an arrow in Fig.~\ref{a2f07}(c), one may observe a large peak
with in-plane character, however,
there is no significant $\alpha^2 \F^{ \overline{xy}, \overline{xy}}$ value at the same location in Fig.~\ref{a2f07}(a).

Now, we look at the projected $\alpha^2\F$ for $\delta E_{\rm F}=-0.205$ eV. As shown in Fig.~\ref{a2f07}(b), the total
$\alpha^2\F$ decreases for $\delta E_{\rm F}=-0.205$ eV comparing with $\delta E_{\rm F}=-0.055$ eV.
However, for the optical modes (modes with energy bigger that 45 meV), and for
$\delta E_{\rm F}=-0.205$, the $\alpha^2 \F^{ \overline{xy}, \overline{xy}}$
corresponding to the phonons with in-plane displacements is enhanced in comparison with that for $\delta E_{\rm F}=-0.055$ eV.
This is true in particular for the lower edge of the optical phonon spectrum marked by an arrow in Fig.~\ref{a2f07}(c), as there is a large peak composed of the phonons with in-plane displacement character, there is a
 sizable $\alpha^2 \F^{ \overline{xy}, \overline{xy}}$ for $\delta E_{\rm F}=-0.205$
at the same location in phonon energy axis in comparison with $\alpha^2\F^{ \overline{xy}, \overline{xy}}$ for $\delta E_{\rm F}=-0.055$ eV. This feature could be partially attributed to the projected DOS in Fig.~\ref{pdos},
where the total DOS is projected into $p_x+p_y$ and $p_z$ orbitals. By moving the $E_{\rm F}$ into the valence states the contribution of states with $p_z$ character decreases and the contribution of states with $p_x+p_y$ character increases. While the former results in reduced the coupling of electronic states with the phonons with major out-of-plane character, the latter results in an enhanced coupling of the electronic states to the phonons with in-plane character.
However, the $\alpha^2\F^{ \overline{xy}, \overline{xy}}$ behaves differently for the phonon energies below 30 meV. In this case the $\alpha^2\F^{ \overline{xy}, \overline{xy}}$ of the $\delta E_{\rm F}=-0.205$ is even smaller than that of $\delta E_{\rm F}=-0.055$. Hence, the argument regarding the enhancement of $\alpha^2\F^{ \overline{xy}, \overline{xy}}$ for $\delta E_{\rm F}=-0.205$ and for phonon energies larger than 45 meV does not hold for phonon energies less than 30 meV.

\begin{figure}[h]%
  \includegraphics*[width=0.4 \textwidth]{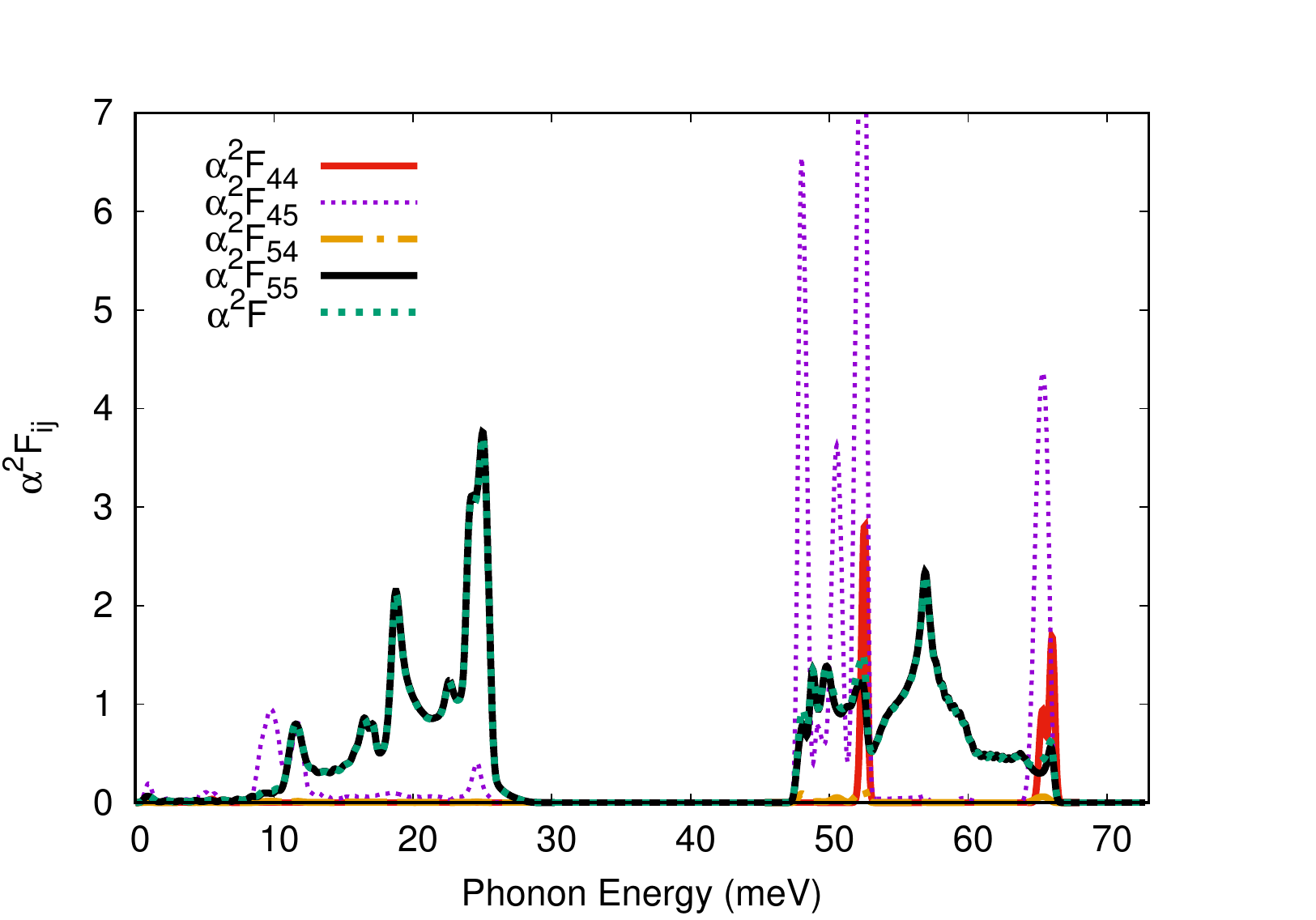}
  \caption[]{%
    (Color online) Band projected $\alpha^2 \F$ for $200\times 200$ $k-$ and $q-$ meshes with Gaussian broadenings $\sigma=0.01$ eV and $\delta{E_{\rm F}}=-0.205$ eV. Notice that band 5 has a major contribution in the $\alpha^2 \F$.}
    \label{a2fprojbnd}
\end{figure}

We have discussed total $\alpha^2 \F$ so far and we have not considered the band anisotropy, in case when we increase $|\delta E_{\rm F}|$ into the valence band states,
the Fermi energy intersects with more than one energy band. To clarify the effects of band anisotropy, we plot the projection of the $\alpha^2 \F$
in Fig.~\ref{a2fprojbnd} for two different bands at the largest examined Fermi shift, $\delta E_{\rm F}=-0.205$ eV. The bands are marked by number 4 and 5 and are labeled by red and green colors, respectively,
in the band structure shown in Fig.~\ref{dosband}.
As seen in Fig.~\ref{a2fprojbnd}, the $\alpha^2 \F_{55}$ (related to band 5) is almost identical to that of
the total $\alpha^2 \F$.
Furthermore, the $\alpha^2 \F_{44}$ (related to band 4) is
very small in comparison to $\alpha^2 \F_{55}$, where the corresponding projected $\lambda$ reads $\lambda_{55}=2.37$ and $\lambda_{44}=0.12$. This could be further understood by noticing that $N_4(0) \ll N_5(0)$. The smallness of the
 $\alpha^2{\bf F}_{44}$ is even more pronounced for a smaller $|\delta E_{\rm F}|$ due to the vanishing $N_4(0)$ (not shown here).
 Therefore, in particular for the actual calculations regarding
  estimation of $T_c$, we only consider band 5 and we neglect the effects of band 4 and its coupling to band 5.
%

\begin{figure}[h]%
  \includegraphics*[width=0.4 \textwidth]{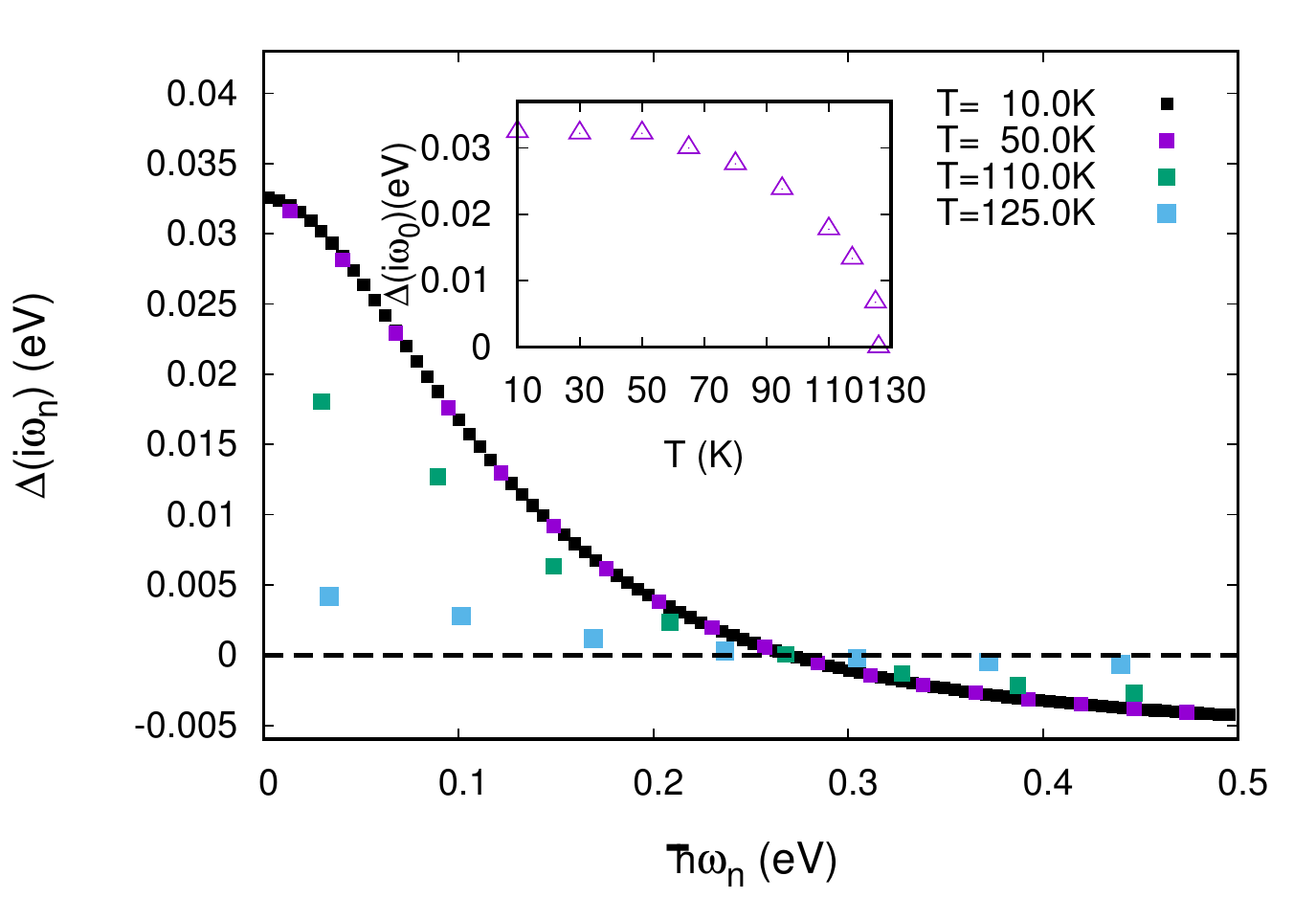}
  \includegraphics*[width=0.4 \textwidth]{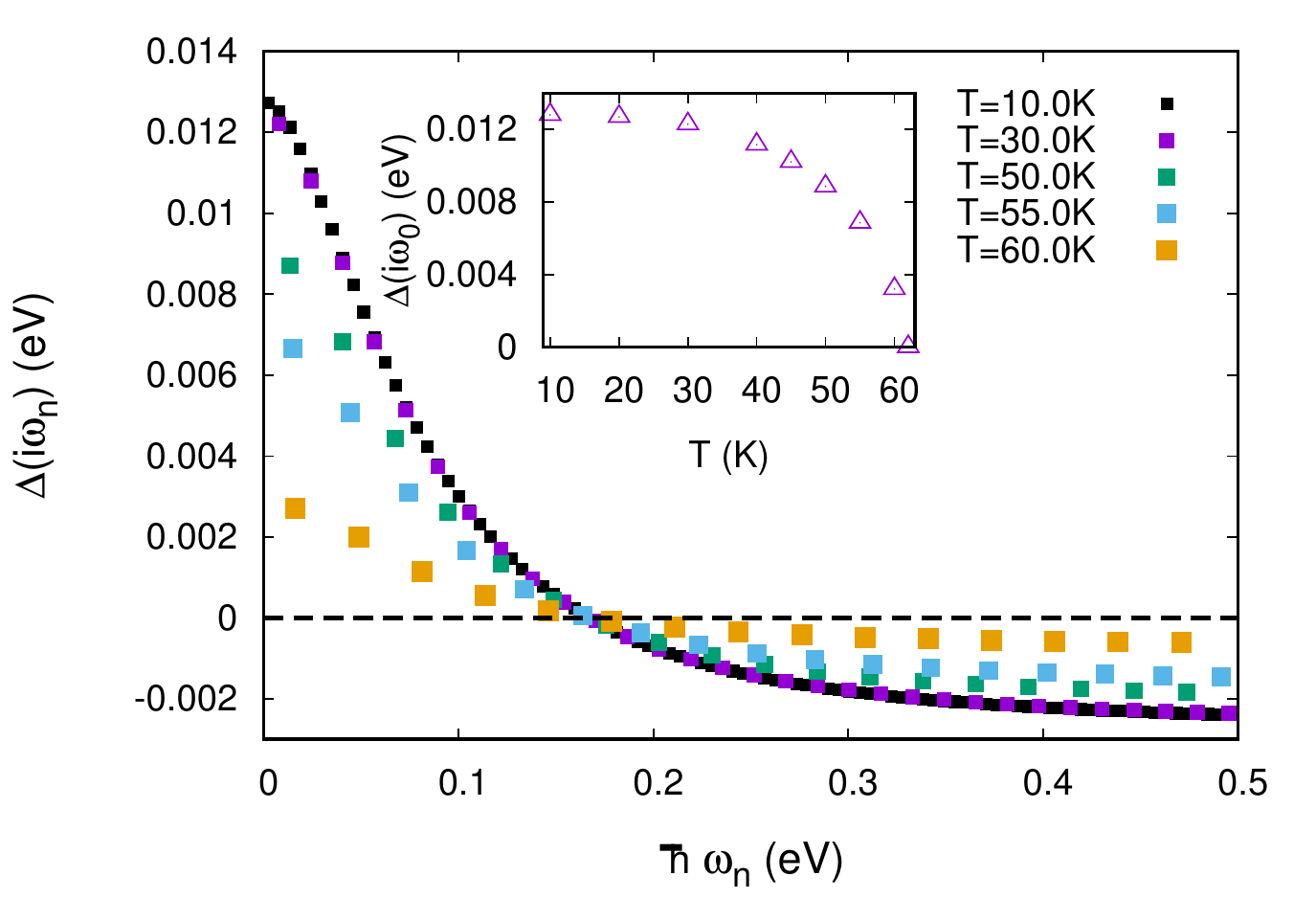}
  \caption[]{%
    (Color online) Superconducting energy gap within a constant DOS approximation for different temperatures with $\mu^*_c=0.1$ at (top) $\delta E_{\rm F}=-0.02$
    and (bottom) $\delta E_{\rm F}=-0.205$ eV in the first-order self-energy approximation. The dashed-dotted line is given as a guide to the eye to determine $T_C$.}
    \label{idelta}
\end{figure}

\begin{figure}[h]%
  \includegraphics*[width=0.4 \textwidth]{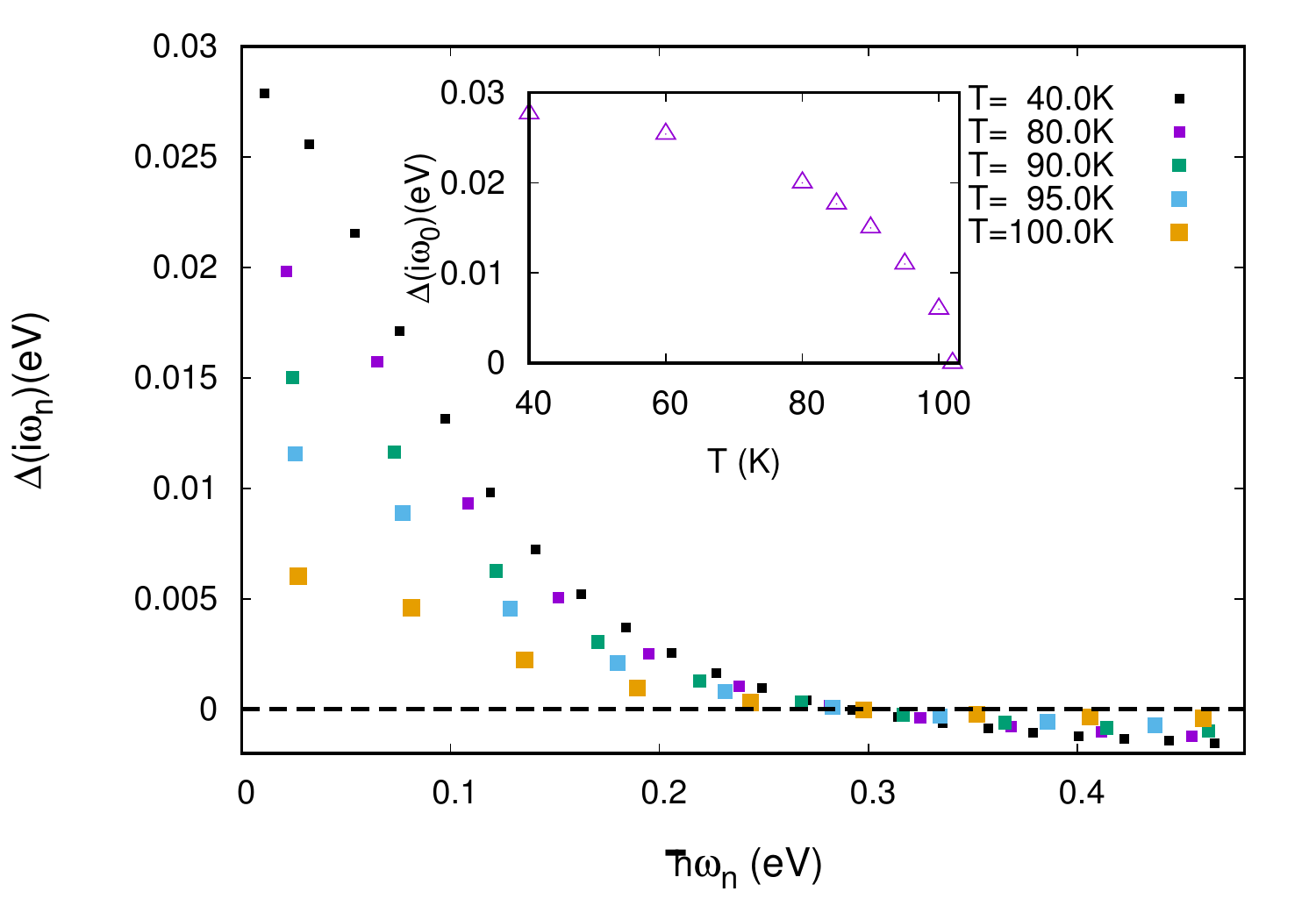}
   \includegraphics*[width=0.4 \textwidth]{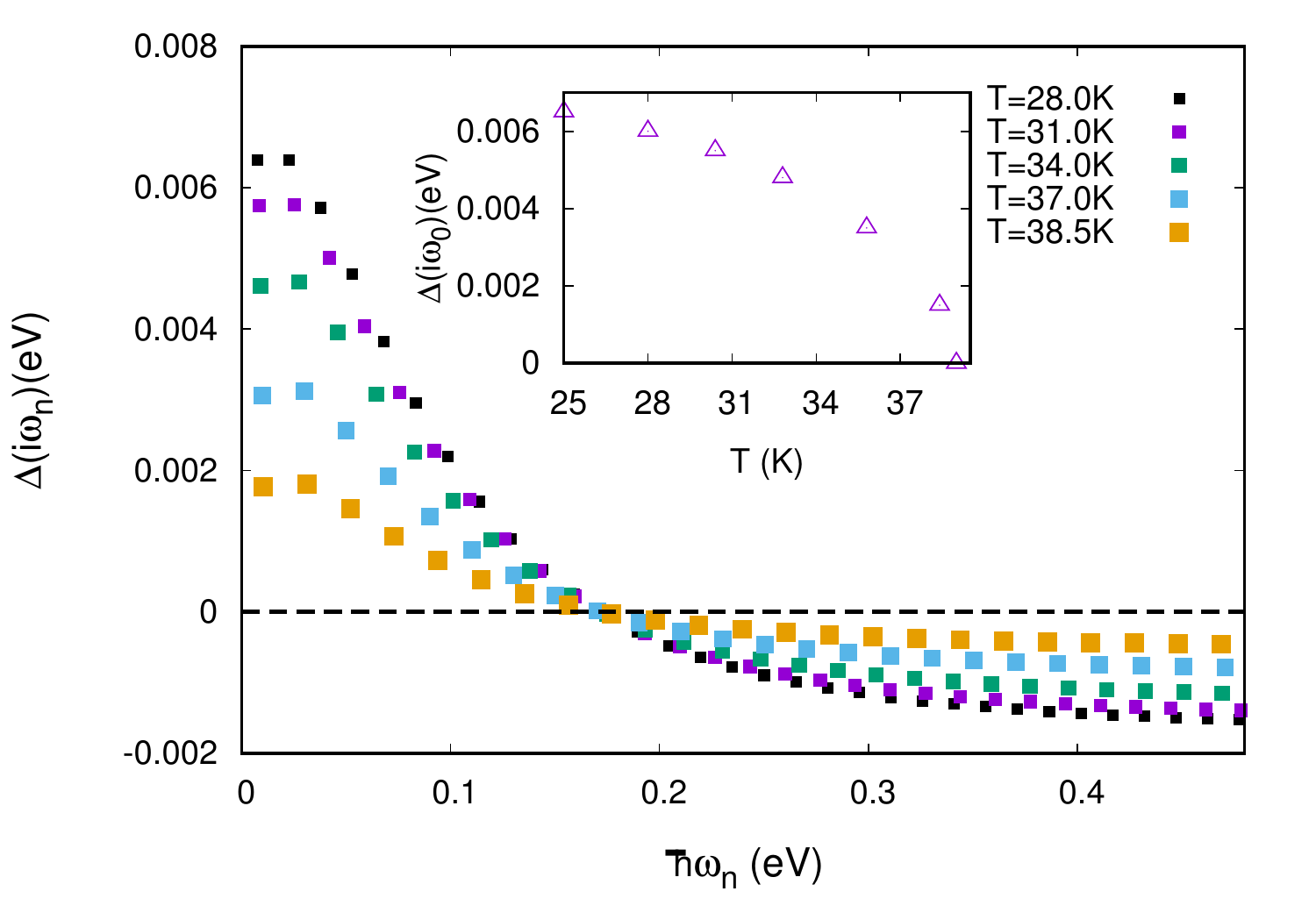}
  \caption[]{%
   (Color online)  Superconducting energy gap, including the vertex corrections and variable DOS approximation for
   different temperatures with $\mu^*_c=0.1$ at (top) $\delta{E_{\rm F}}=-0.02$ and (bottom) $\delta{E_{\rm F}}=-0.205$ eV. The dashed-dotted line is given as a guide to the eye to determine $T_C$. }
    \label{ideltavertex}
\end{figure}

In order to calculate the energy gap, we do need to calculate the Green's function on the real frequency axis using the analytical
continuation~\cite{vidberg,leavans,mars-analyt}. The superconducting energy gap can be defined as the
energy difference between the ground state of the superconductor and the energy of the lowest quasi-particle excitation~\cite{tinkham}.
Furthermore, the effective energy gap in superconductors can be measured in
microwave absorption experiments.

We are just interested in the critical temperature,
which is also obtained by the zeroth of the energy gap along the imaginary frequency. By using the self-energy decomposition, the gap function,
 $\Delta(i\omega_n)$ is defined as $\Delta(i\omega_n) = \phi(i\omega_n)/Z(i\omega_n)=\Sig_{12}(i\omega_n)/Z(i\omega_n)$.

Having calculated the $\alpha^2 \F_{55}$ which is almost identical to total $\alpha^2{\bf F}$ for
all examined $\delta E_{\rm F}$ (for example see Fig.~\ref{a2fprojbnd},
we solve the isotropic
Eliashberg equations for different values of the Fermi energy shifts,
namely, $\delta E_{\rm F}=-0.02, -0.055, -0.105, -0.155$ and $-0.205$ eV.

In the following we solve Eliashberg equations within different approaches discussed in the text and are compared to each other.
 We consider, namely Migdal-Eliashberg $+$ constant DOS approximation (Eqs. (\ref{eliashbergsingle-muc}), called ConsDOS),
 Migdal-Eliashberg $+$ variable DOS approximation (Eqs.~(\ref{s1-od-muc}), (\ref{dyson}) and (\ref{g022}) called VarDOS) and
 VarDOS $+$ vertex corrections (the second-order diagram is included through Eqs.~(\ref{ss21a}-\ref{ss21}), Eqs.~(\ref{dyson}) and (\ref{g022}) called Vertex).
 For the sake of completeness, we estimate the $T_c$ using the Allen-Dyns modified
 McMillan equation~\cite{allen}, where the critical temperature is given by
 $T_c=\frac{\omega_{in}}{1.2}\exp[-\frac{1.04(1+\lambda)}{\lambda-\mu^*_c(1+0.62\lambda)}]$ where $\omega_{in}$ is
 the logarithmical averaged frequency. Hereafter, we set $\mu^*_c=0.1$.
 Vanishing the gap function $\Delta(i \omega)$ is the criteria for finding the
 $T_c$.

In Fig.~\ref{idelta} we illustrate the variations of the $\Delta(i\omega)$ for different temperatures and
for two different shifts at $\delta E_{\rm F}=-0.02$ and $-0.205$ eV within ConsDOS approach.
Furthermore, in Fig.~\ref{ideltavertex} the same plot is shown for $\delta E_{\rm F}=-0.02$ and $-0.205$ eV,
by employing Vertex approach.
The reduction of the energy gap as temperature attain to the critical temperature can be taken as an indication that the charge carriers have a kind of a collective nature.
That is, the charge carriers must consist of at least two things which are bound together, and the binding energy is weakening as temperature attain the critical temperature.
Above the critical temperature, such collections do not exist, and normal resistivity prevails.

\begin{figure}[h]%
  \includegraphics*[width=0.4 \textwidth]{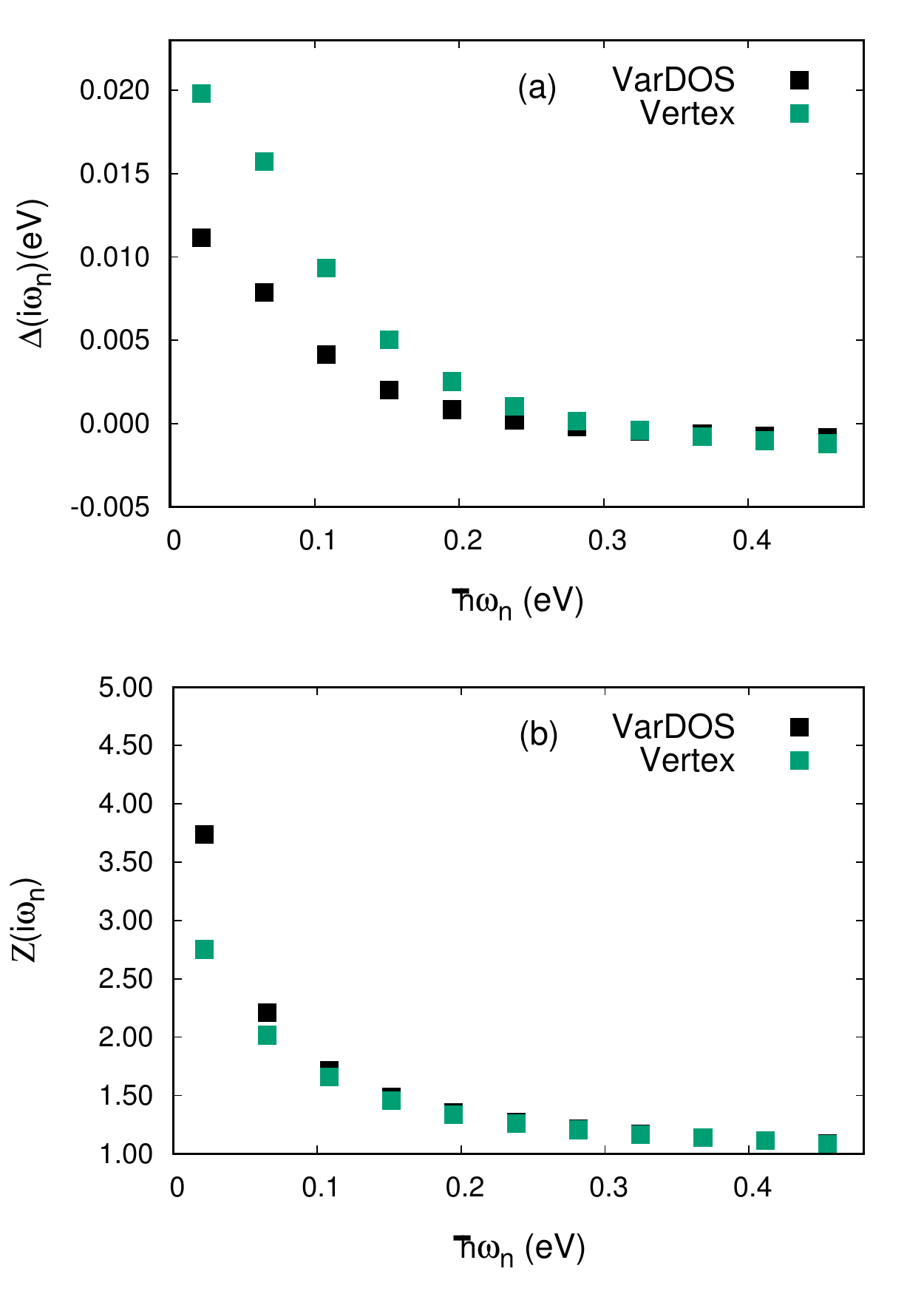}
  \caption[]{%
    (Color online) (a) Superconducting energy gap and (b) mass renormalization function for $\delta E_{\rm F}=-0.02$ eV
    and $T=80$ K comparing two approaches, VarDOS and Vertex approximations.
      }
    \label{ideltaizT80}
\end{figure}

The calculated $T_c$ for different $\delta E_{\rm F}$ is presented in
Table~\ref{tab1} which is our main results in this paper.
Furthermore, in Table~\ref{tab1} the $T_c$ is estimated and compared by using different approaches mentioned before.
As seen in the Table~\ref{tab1}, generally speaking, while Allen-Dynes approach acquires the smallest estimate of
the value of the $T_c$ for all cases, the ConsDOS approximation overestimates the value of $T_c$
in comparison with the VarDOS approach. The discrepancy between the ConstDOS and VarDOS is larger for $\delta E_{\rm F}$
which are located in the proximity to the apex of the DOS, where the variations of the DOS is large. For a larger shift of $E_{\rm F} $ to lower energies,
the disagreement between ConsDOS and VarDOS approximations
becomes smaller. The role of the vertex correction over VarDOS is non-trivially depends on the structure of
the energy dispersion around the Fermi energy. Apparently, the vertex corrections are constructive to the value of $T_c$
for $E_{\rm F}$ near the peak of the DOS, namely $\delta E_{\rm F}=-0.02$ and $-0.055$ eV. However, the vertex correction
is detrimental to $T_c$ for the shifts away from the peak of the DOS.
\begin{table}[h]
\begin{tabular}{l*{7}{c}r}
\hline\hline
\vline~$\delta E_{\text F}$  & $T_c(\text{Vertex})$ & $T_c(\text{VarDOS})$ & $T_c(\text{ConstDOS})$& Allen-Dynes &\vline \\
\hline
\vline~-0.02 & 103 & 88 & 126& 62 &\vline   \\
\hline
\vline~-0.055            & 98 & 88 & 131&  64 &\vline \\
\hline
\vline~-0.105           & 62 & 82 & 89& 56 &\vline \\
\hline
\vline~-0.155     & 43 & 72 & 72& 50 &\vline \\
\hline
\vline~-0.205     & 39 & 60 & 61& 45 &\vline \\
\hline\hline
\end{tabular}
 \caption[]{%
 Superconductive critical temperature, $T_C$, in units of Kelvin for different $E_{\rm F}$
       shifts and approximations. Notice, we set $\mu^*_c=0.1$ and the vertex
  corrections are included only for the electron-phonon part. The corresponding hole densities
  for mentioning $\delta E_{\text F}$ are $5.0\times10^{13}$, $1.5\times10^{14} $, $2.4\times10^{14}$
  , $3.1\times 10^{14}$ and $3.8\times10^{14}$ cm$^{-2}$, respectively.}
  \label{tab1}
\end{table}
To further explore the effect of the vertex correction, for $\delta E_{\rm F}=-0.02$ eV, we plot in Fig.~\ref{ideltaizT80}(a) and (b)
the superconducting gap and mass renormalization at $T=80 K$ respectively,
where the VarDOS solution is compared with that based on the Vertex approach.
  Obviously, the superconducting gap for the vertex corrected one is larger than that obtained in the VarDOS solution, however, the Vertex
  approach acquires smaller mass renormalization in comparison with that calculated in the VarDOS approach. This observation, is the signature of achieving
    larger $T_c$ with smaller effective $\lambda$ through inclusion of vertex corrections~\cite{grim}.

\section{Conclusion}~\label{sec:conc}
In conclusion, we have investigated a possible superconductivity of hole doped BLP. Owing to the mutual presence of nearly flat band near the VBM together with
a breaking of $\sigma_h$ symmetry in BLP, a larger electron-phonon interaction appears, upon the hole doping system.
By projecting the $\alpha^2{\bf F}$ into different phonon deformations, it is revealed that the out-of-plane displacement of the
phonons have the largest contribution to electron-phonon interactions. For optical phonons, the contribution of the in-plane
displacements are increased upon furthering $\delta E_{\rm F}$ into VBM states,
where the electronic band structure near the Fermi surface acquires larger $p_x+p_y$ character, leading to an enhanced coupling of
the electrons to the modes with larger in-plane displacements.
 We have further examined different approaches calculating $T_c$ of the BLP for different shifts of $E_{\rm F}$. The brief description of the approaches we have used is as follows: (i) Allen-Dynes formula,
 (ii) the first-order self-energy diagram (Migdal-Eliashberg) within assumption of a constant DOS at $E_{\rm F}$ (called ConstDOS) (iv) Considering the first-order self-energy diagram
 (Migdal-Eliashberg) by solving full DOS variations (called VarDOS) and (vi) assuming the ${\bf k}$ averaged second-order diagram over the VarDOS approach (called Vertex).

 Our calculations, summarized in Table I, show that for all $\delta E_{\rm F}$, Allen-Dynes formula estimates smaller $T_c$ in comparison with the other approaches.
 The ConstDOS overestimates $T_c$ in comparison to the VarDos approach which the disagreement between the two approaches appear to be smaller for a larger $\delta E_{\rm F}$ below VBM.
 The effect of the Vertex differs based on the $\delta E_{\rm F}$. While the vertex correction enhances $T_c$ for $\delta E_{\rm F}=-0.02$ and $-0.055$ eV, it is detrimental to a larger examined $\delta E_{\rm F}$.

We have shown in this work that the high superconducting critical temperature occurs for a hole doped blue phosphorene ranging from $100$ to $40$K by considering the hole
densities between $5\times10^{13}$ to $3.8\times 10^{14}$ cm$^{-2}$ and our prediction should be verified by current experiments.
\begin{figure}[h]%
	\includegraphics*[width=0.4 \textwidth]{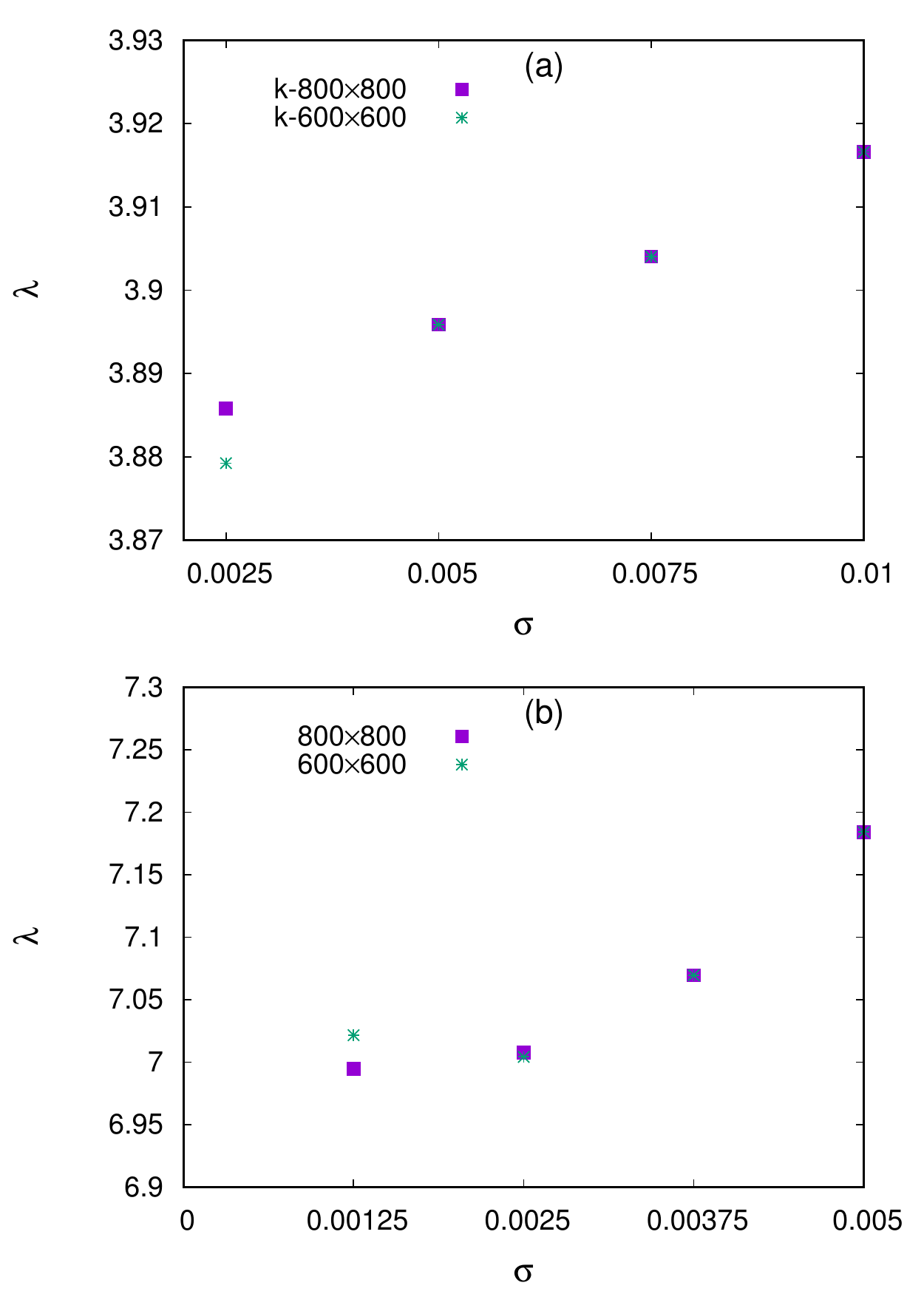}
	\caption[]{%
		(Color online)  (a) Total unit-less electron-phonon coupling $\lambda$ for $\delta E_{\rm F}=-0.055$ eV as a function of Gaussian
		broadening $\sigma$ for
		two different fine $k-$mesh. (b) the same as plot (a) but for $\delta E_{\rm F}=-0.105$ eV.}
	\label{figappb1}
\end{figure}
\begin{figure}[h]%
	\includegraphics*[width=0.4 \textwidth]{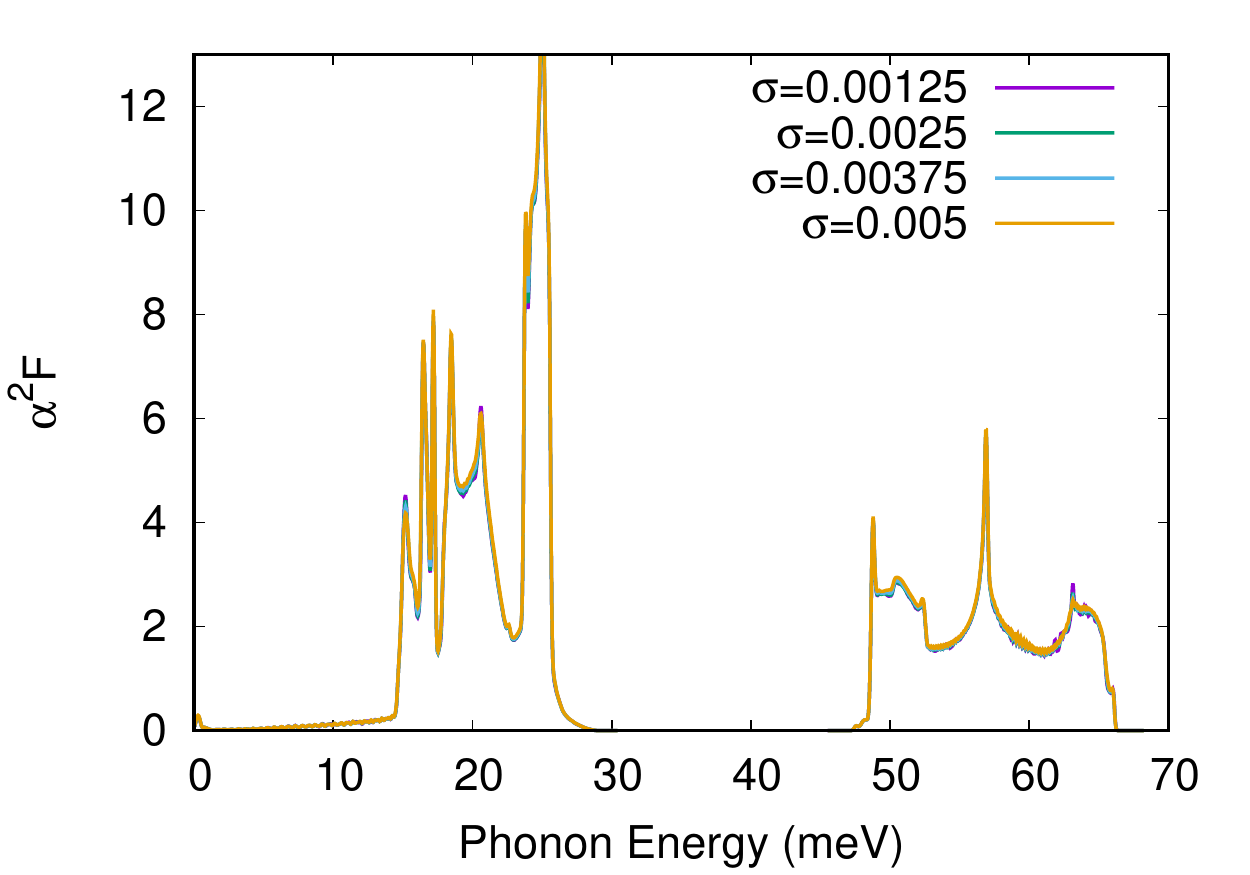}
	\caption[]{%
		(Color online) (a) Total $\alpha^2{\bf F}$ for $800\times 800$ fine k- and $200\times 200$  fine $q-$ meshes as a function of electronic
		Gaussian broadening $\sigma$ for $\sigma_{ph}=0.1$ and $\delta E_{\rm F}=-0.055$ eV.}
	\label{figappb2}
\end{figure}

\section{Acknowledgement}
We would like to thank M. Vozmediano, F. Guinea and D. Daghero for fruitful discussions. This work is partially supported by the Iran Science Elites Federation grant.

\appendix
\section{Appendix}~\label{sec:appa}
The electron-phonon matrix elements are defined as,
\begin{equation}
g^{\nu\sigma}_{\k i,\k' j} =\left({\hbar\over 2\omega_{\q,\nu}}\right)^{1/2}
\langle\psi_{\k i \sigma}| {\Delta^{\q\nu}V_{\bf KS} }|\psi_{\k' j \sigma}\rangle.
\end{equation}
with $\q=\k'-\k$, $\displaystyle \Delta^{\q\nu}V_{\bf KS} = \sum_{s\kappa} \frac{\partial V_{\bf KS}}{\partial {\bf u}^{\bf q}_{s\kappa}} {\bf u}^{{\bf q}\nu}_{s\kappa}$ and $\bf k$ is an electron wave vector,
$\nu$ is the index of the phonon mode which contributes to the scattering of the electrons, $\bf q$ is the phonon wave vector, ${\Delta^{\q\nu}V_{\bf KS} }$ is the potential owing to the displacement pattern of the phonon mode $\nu$ and
$\frac{\partial V_{\bf KS}}{\partial {\bf u}^{\bf q}_{s\kappa}}$ is the potential difference due to a displacement of ${\bf u}^{\bf q}_{s\kappa}$.  Here, $s$ is the index of atoms in the unit-cell,
$\kappa=\bar{x},\bar{y},\bar{z}$ is the Cartesian direction index, displacement vector ${\bf u}^{{\bf q}\nu}_{s}$ is mass renormalized polarization vector, i.e
${\bf u}^{{\bf q}\nu}_{s} =\frac{1}{\sqrt{M_s}}{\bf e}^{{\bf q}\nu}_{s}$, where vector ${\bf e^{\q\nu}}$ is the eigenvector of the dynamical matrix~\cite{baroni}. For the illustrative purposes, we define a Cartesian projected electron-phonon coupling as,
\begin{equation}
g^{\nu,\kappa}_{\k i,\k' j} = \left({\hbar\over 2\omega_{\q,\nu}}\right)^{1/2}
\langle\psi_{i\k}| {\Delta^{\q\nu}_\kappa V_{\bf KS} }|\psi_{j,\k'}\rangle.
\end{equation}
with $\displaystyle \Delta^{\q\nu}_\kappa V_{\bf KS} =\sum_{s} \frac{\partial V_{\bf KS}}{\partial {\bf u}^{\bf q}_{s\kappa}} {\bf u}^{{\bf q}\nu}_{s\kappa}$ and $g^{\nu,\kappa}_{\k i,\k' j}$
which satisfies $g^{\nu}_{\k i,\k' j}=\sum_\kappa g^{\nu,\kappa}_{\k i,\k' j}$. \\

\section{Appendix}~\label{sec:appb}
The different quantities i. e. DOS, $\alpha^2{\bf F}$ and $\lambda$ depend on
electronic mesh size $N_k$ and electronic Gaussian broadening $\sigma$. We are interested in the limit
$N_k \rightarrow \infty$ and $\sigma \rightarrow 0$ and this is also true for phononic $q-$mesh size and phononic Gaussian
broadening $\sigma_{ph}$ as well.
Due to the presence of a double delta summation over $k-$mesh in the evaluation of $\lambda$ and $\alpha^2{\bf F}$, the convergence
of $\lambda$ and $\alpha^2{\bf F}$
as a function of $\sigma$ and $N_k \times N_k$ is difficult. Therefore, by using a $200\times 200$ $q-$mesh and $\sigma_{ph}=0.1$ meV,
we show in which range of $N_k$ and $\sigma$ the above mentioned quantities are insensitive to the value of
the $N_k$ and $\sigma$. To this end, in Fig.~\ref{figappb1} the total unit-less electron-phonon coupling $\lambda$
is depicted as function of $N_k$ and $\sigma$. The numerical result shows more fluctuation as a function of $\sigma$ for $\delta E_{ \rm F}=-0.055$ eV due to its vicinity to the peak of the DOS. Therefore, results are more stable for the smaller value of $\sigma$.
Moreover, it is obvious that for the applied range of $\sigma$ shown in Fig.~\ref{figappb1}, the $\lambda$ is almost converged as a function of mesh-size.

In Fig.~\ref{figappb2}, total $\alpha^2{\rm F}$ is shown for different values of $\sigma$ at $\delta E_{\rm F}=0.055$ eV. As it is clear from the figure, $\alpha^2{\rm F}$ is almost insensitive to the $\sigma$.

\end{document}